\documentclass[pra,aps,showpacs,superscriptaddress,twocolumn,10pt]{revtex4-1}
\usepackage{hyperref,braket}
\usepackage{braket}
\usepackage[cm]{fullpage}
\usepackage{graphicx}	
\usepackage[english]{babel}
\usepackage{amsmath}
\usepackage{amssymb,bm}
\usepackage{cancel}
\usepackage{natbib}
\usepackage{comment}
\usepackage{xcolor}
\usepackage[normalem]{ulem}
\usepackage{bbold}
\usepackage[outdir=./]{epstopdf}
\usepackage{here}

\usepackage{mathtools}

\newcommand{\Nit}{{\cal N}}
\DeclareMathOperator*{\argmax}{arg\,max}
\DeclareMathOperator{\tr}{tr}
\DeclareMathOperator{\ig}{IG}
\DeclareMathOperator{\qcb}{QCB}
\DeclareMathOperator{\prob}{P}
\usepackage{makecell}

\newcommand{\xt}{\bf{x}_{\textrm{true}}}

\newcommand{\pe}{P_{\textrm{e}}}
\newcommand{\pen}{P_{\textrm{e},N}}
\newcommand{\pencl}{P^{\textrm{cl}}_{\textrm{e},N}}
\newcommand{\rhoin}{\rho_{\textrm{in}}}
\newcommand{\thetat}{\theta_{\textrm{true}}}
\newcommand{\gammat}{\gamma_{\textrm{true}}}

\makeatletter
	\renewcommand{\maketag@@@}[1]{\hbox{\m@th\normalsize\normalfont#1}}
\makeatother
\begin{document}

\author{Nadia Milazzo}
\affiliation{Institut f\"ur theoretische Physik, Universit\"at T\"{u}bingen, 72076 T\"ubingen, Germany}
\affiliation{Université Paris Saclay, CNRS, LPTMS, Orsay 91405, France}
\affiliation{{ColibrITD, 91, rue du Faubourg Saint-Honor\'e
75008 Paris }}
\author{Olivier Giraud}
\affiliation{Université Paris-Saclay, CNRS, LPTMS, 91405 Orsay, France}
\author{Giovanni Gramegna}
\affiliation{Institut f\"ur theoretische Physik, Universit\"at T\"{u}bingen, 72076 T\"ubingen, Germany}
\author{Daniel Braun}
\affiliation{Institut f\"ur theoretische Physik, Universit\"at T\"{u}bingen, 72076 T\"ubingen, Germany}

\begin{abstract}
  With increasing commercial availability of quantum information
  processing devices the need for testing them efficiently for their
  specified functionality will arise.  Complete quantum channel
  characterization is out of the question for anything more than the
  simplest quantum channels for one or two qubits. Quantum functional
  testing leads to a decision problem with outcomes corresponding to
  rejection or acceptance of the claim of the producer that the device
  parameters are within certain specifications.  In this context, we introduce
  and analyse three ingredients that can speed up this decision problem: iteration of the channel, efficient decision criteria, and non-greedy adaptive experimental design. 
\end{abstract}

\title{Principles of quantum functional testing}

\date{\today}

\maketitle

\section{Introduction}
Functional testing in the semi-conductor context refers, broadly, to testing whether the device under test performs as designed, and detecting and possibly preparing repair of faulty behavior (e.g.~by activating spare parts of a chip).  With increasing integration density, functional testing has become one of the most expensive and time-consuming parts of chip production. With the number of memory cells on a DRAM chip in the billions, it is clearly out of the question to test all possible states of the memory as the time needed would by far exceed the life-time of the Universe.  Rather, test engineers focus on specific parts of the chip, such as address decoders, actual memory cells, sense-amps etc., and known error mechanisms. The latter typically include interaction of a cell with neighboring cells. E.g.~flipping one memory cell might flip a physically neighboring cell through various mechanisms.  Many such error mechanisms can be tested in parallel by writing and reading out specific patterns into the memory, such as checkerboard patterns of 0s and 1s.  \\

While there is no mass production of quantum-information-processing devices yet, it is expected that sooner or later similar considerations of quantum functional testing will be needed.  Clearly, due to the essential necessity of quantum-coherent superpositions, quantum functional testing is an even more daunting task than classical functional testing.  
Several methods have been proposed for characterizing quantum channels as well as possible.  Most focus has been on ``quantum channel tomography'' \cite{chuang_prescription_1997,PhysRevA.63.020101,PhysRevLett.86.4195,PhysRevLett.90.193601,PhysRevA.64.012314,leonhardt_tomographic_1995,kosut_quantum_2008,christandl_reliable_2012,blume-kohout_robust_2012,rey-de-castro_time-resolved_2013,ohliger_efficient_2013}, with prominent methods such as randomized benchmarking 
\cite{knill_randomized_2008,bendersky_selective_2009,cramer_efficient_2010,magesan_scalable_2011,magesan_efficient_2012,gaebler_randomized_2012,onorati_randomized_2018,cross_validating_2018,nielsen_gate_2021}, cross-entropy benchmarking \cite{boixo_characterizing_2018}  (the latter was used in \cite{Arute2019} for showing ``quantum supremacy''), Bayesian techniques 
\cite{huszar_adaptive_2012,huszar_adaptive_2012,ferrie_self-guided_2014,kueng_near-optimal_2015,granade_practical_2016}, compressed sensing \cite{gross_quantum_2010,flammia_quantum_2012}, matrix-product state representations \cite{baumgratz_scalable_2013,lanyon_efficient_2016}, multiplexed measurements \cite{titchener_scalable_2018}, or machine learning 
\cite{Fiderer2021,quek_adaptive_2021,torlai_neural-network_2018};  see also the reviews and analyses in \cite{mohseni_quantum-process_2008,banaszek_focus_2013,schwemmer_systematic_2015}. Recently, for NISQ devices the fitting and extrapolation of error models \cite{Arute2019}, and the study of correlations of errors \cite{PRXQuantum.1.010305} have started to receive increased attention. 

Functional testing takes a perspective that differs from quantum channel tomography. At the heart is a 
decision problem: the device should be accepted or rejected (or
possibly qualified as repairable, with an additional output of what repair is needed), and since testing time is expensive, this decision must be reached as quickly as possible. In particular, the test engineer wants to waste as little time as possible on devices that do {\em not} function properly.  In addition, it is desirable to provide a level of confidence that the decision is correct, and tests continue until such a level of confidence for a decision is reached, or testing is stopped if a specified maximum number of tests have been performed without reaching the required level of confidence. Naturally then, new tools from statistical analysis, notably the Chernoff bound \cite{Chernoff1952} and its quantum generalization, the quantum Chernoff bound \cite{Audenaert2007,Audenaert2008,Nussbaum2009}, become important.  They provide asymptotic error rates for discriminating classical probability distributions, or quantum states \cite{pirandola_fundamental_2019},  respectively.  

The world of classical functional testing divides into analogue and digital testing.  In the former falls testing of e.g.~sense amplifiers, current drivers etc.~that can be characterized by a set of continous parameters, in the latter testing of Boolean functionality (including the identity function for memories).  Quantum devices, even quantum memories, fall naturally and at least partly in the realm of  analogue testing, due to the fact that quantum states vary continuously, in spite of the fact that in the context of quantum error correction errors can be ``quantized''.    E.g.~a single-qubit memory cell should be able to store and retrieve any linear superposition of $\ket{0}$ and $\ket{1}$, parametrized by two angles of the Bloch vector, and a quantum channel is typically characterized by a set of continuous parameters.  Fundamentally, the decision problem should therefore be based on quantum parameter estimation and has similarity with quantum metrology --- the difference being, however, that in the latter one strives at as high sensitivity as possible, whereas in quantum functional testing one strives at arriving at estimates with large confidence as quickly as possible.  Crucial for this is the availability of a ``spec'', i.e.~a specification of the device provided by its producer (called ``the company'' in the following).  

The specific quantum-mechanical context opens room for new principles for functional testing. In this paper we introduce and investigate three such principles: iteration of the channel, efficient decision criteria,  and non-greedy Bayesian updates. In Sec.~\ref{sec.it} we discuss the relevance of iterating the channel, and in Sec.~\ref{ngu} we present the Bayesian context in which our protocols are studied. In Sec.~\ref{applications} we apply these principles to the problem of certifying quantum channels.  
Ideas in the direction of a quantum-functional-testing approach have recently also been put forward under the label ``quantum system certification'' \cite{Kliesch2021}.

\section{Iterating a quantum channel}\label{sec.it}

\subsection{Intuition: coherent enhancement of errors \label{motiv}}
Any unitary single-qubit channel $\Phi:\mathbb C^2\to \mathbb C^2$ can be written as $\rho\mapsto \rho'=\Phi(\rho)=R_{\hat{\bm n}}(\theta)\rho R_{\hat{\bm n}}^\dagger(\theta)$ with 
\begin{equation}
  \label{eq:ch}
R_{\hat{\bm n}}(\theta)=e^{-i \frac{\theta}{2} \hat{\bm n}\cdot\bm\sigma}\,, 
\end{equation}
where $\hat{\bm n}$ is a unit vector about which the Bloch vector of the state is rotated by an angle $\theta$, and $\bm\sigma$ is the vector of Pauli matrices. To see why iterations of a channel might be beneficial for speeding up the decision problem, consider a phase gate with $R_{{\bm e}_z}(\theta)$, i.e.~a rotation about the $z$-axis, and an initial state $\ket{+}$, with $\ket{\pm}=(\ket{0}\pm\ket{1})/\sqrt{2}$. After application of the channel, one measures ideally in a basis $\{R_{{\bm e}_z}(\theta)\ket{+},R_{{\bm e}_z}(\theta)\ket{-}\}$, which will lead to probabilities 1 and 0, respectively, to find $+$ and $-$. Now suppose that the gate actually enacts the (fixed) faulty unitary $R_{{\bm e}_z}(\theta')$ with $\theta'=\theta+\varepsilon$, where $\varepsilon$ is a small angle. The probability to find $+$ and $-$ in the same measurement are then, respectively, $\simeq 1-\varepsilon^2/4$ and $\simeq\varepsilon^2/4$. The chance to detect the faulty behavior can be enhanced by sampling the same channel $N_0$ times, in which case the probability to find a ``$-$'' increases to $\simeq N_0\varepsilon^2/4$. But a much more efficient strategy is to iterate the channel $\Nit$ times and only measure at the end in the correspondingly adapted basis $\{R_{{\bm e}_z}(\Nit\theta)\ket{+},R_{{\bm e}_z}(\Nit\theta)\ket{-}\}$, in which case the faulty channel will lead to a probability for the outcome $''-''$ given by $(\Nit\varepsilon/2)^2$ (for $\Nit\varepsilon\ll 1$).  For fair comparison, and supposing that the application of the channel dominates the total time required for the experiment rather than the measurement itself, we can sample the output after $\Nit$ iterations $N=N_0/\Nit$ times.  The probability to find an outcome ``$-$'' will then be $\simeq (N_0/\Nit)(\Nit\varepsilon/2)^2=N_0 \Nit \varepsilon^2/4$, i.e.~the coherence of the channel can be exploited to gain a factor $\Nit$ in efficiency. 
 Iterations of channels have been proposed previously in the context of ``long sequence gate set tomography'' \cite{nielsen_gate_2021}. In the remainder of this section we study in more detail under what conditions iterations of the channel are beneficial: If $\varepsilon$ is not so small or is random with a certain probability distribution, or if decoherence effects play a role, then the advantage of iterating is not so clear. In addition, the answer might depend on the measurements available. A more rigorous assessment using a Chernoff bound is then in order; this will be developed in the next subsection. In Sec.~\ref{ngu} we shall propose a setup which automatically determines at each step of the protocol whether to iterate or measure directly.

\subsection{Quantum Chernoff bound}
\label{secchernoff}
The problem of {\em discriminating} two different quantum channels,  such as an ideal channel from a faulty one (as opposed to fully characterizing one of them), can be cast in the framework of quantum hypothesis testing for quantum states (see e.g.~\cite{Audenaert2008}) by calculating the quantum Chernoff bound between the output states returned by the channels considered. Classically, the error probability $\pe$ in discriminating two probability distributions (by betting on the probability distribution that is more likely to generate some observed outcome) decreases exponentially with the number $N$ of draws from the distributions in the limit $N\to\infty$,
\begin{equation}
\pencl\sim e^{-N \xi}\,.
\label{Pe}
\end{equation}
The asymptotic error exponent $\xi$ (called ``Chernoff bound'' in the following) was found by Chernoff in \cite{Chernoff1952} and later generalized to the quantum situation in \cite{Audenaert2007,Nussbaum2009}. In the quantum case we are given $N$ identical copies of a quantum state (or density matrix) that can only be one out of two possible states, $\rho$ or $\tau$, and are supposed to determine which quantum state has been provided. In a Bayesian setting, let $\pi_{0} >0$ and $ \pi_{1} >0$ be the prior probabilities with which $ \rho $ and $ \tau $ occur, respectively. In the absence of any a priori knowledge on $\tau$, we have $\pi_{0}=\pi_{1}=\frac12$. The minimal error probability based on an optimized POVM measurement can be shown \cite{Audenaert2008} to take the form 
\begin{align}
\label{chernoff2}
&\pen= \frac12\left(1-\Vert \pi_{1}\tau^{\otimes N}-\pi_{0}\rho^{\otimes N} \Vert_1\right)\,,
\end{align}
where $ \Vert O\Vert_1=\tr\vert O\vert $ is the trace norm. This error probability decreases again exponentially in the asymptotic limit $N\to\infty$, which allows one to define the \emph{quantum Chernoff bound} $ \xi_{\qcb} $, 
\begin{align}
&\xi_{\qcb} =-\underset{N\rightarrow\infty}{\lim}{\frac{1}{N}}\log \pen\,.
\end{align}
In \cite{Audenaert2008} an analytical formula for $ \xi_{\qcb} $ was found in terms of the two states $\rho$ and $\tau$ we wish to discriminate. It reads
\begin{equation}
\xi_{\qcb}(\rho,\tau) = -\log(\underset{0\leq s \leq 1}{\inf} \tr(\rho^{1-s}\tau^{s}))\,.
\label{chernoff}
\end{equation}

In the context of quantum functional testing, we set $ \tau $ as the output of the ideal channel and $ \rho $ as the output of the faulty channel, for given input state $\rhoin$. The quantum Chernoff bound $\xi_{\qcb}$ then gives the asymptotic rate of decrease of the minimal probability of not discriminating the faulty from the ideal channel upon measurement of the output. In order to maximize the rate at which the outputs are discriminated, we additionally maximize $ \xi_{\qcb} $ over all possible input states $\rhoin$. It turns out that it is sufficient to maximize over pure input states: indeed, denoting with $\rhoin=\sum_i p_i \ket{\psi_i}\!\bra{\psi_i}$ any decomposition of $\rhoin$ in pure states, one has:
\begin{align}\notag
	&\xi_{\qcb} (\Phi_{\mathrm{ideal}}(\rhoin),\Phi_{\mathrm{faulty}}(\rhoin))\\
	\notag
	&\qquad\leqslant \sum_i p_i \xi_{\qcb} (\Phi_{\mathrm{ideal}}(\ket{\psi_i}\!\bra{\psi_i}),\Phi_{\mathrm{faulty}}(\ket{\psi_i}\!\bra{\psi_i}))\\
	&\qquad\leqslant \sup_{i} \xi_{\qcb} (\Phi_{\mathrm{ideal}}(\ket{\psi_i}\!\bra{\psi_i}),\Phi_{\mathrm{faulty}}(\ket{\psi_i}\!\bra{\psi_i})),\label{eq:convexity}
\end{align}
where the first inequality holds by virtue of the joint convexity of $\xi_{\qcb}$
\cite{Calsamiglia2008,Audenaert2008,lieb_convex_1973}  and linearity of quantum channels. 
As a consequence of \eqref{eq:convexity} it is always better to consider pure input states rather than mixed ones. In what follows we will therefore, in the case of qubit channels, optimize over initial pure states parametrized by their Bloch vector $\mathbf{r}_\textrm{in}=(\sin\alpha\cos\beta,\sin\alpha\sin\beta,\cos\alpha)$.

\begin{figure}[!t]
\centering{\includegraphics[scale=0.6]{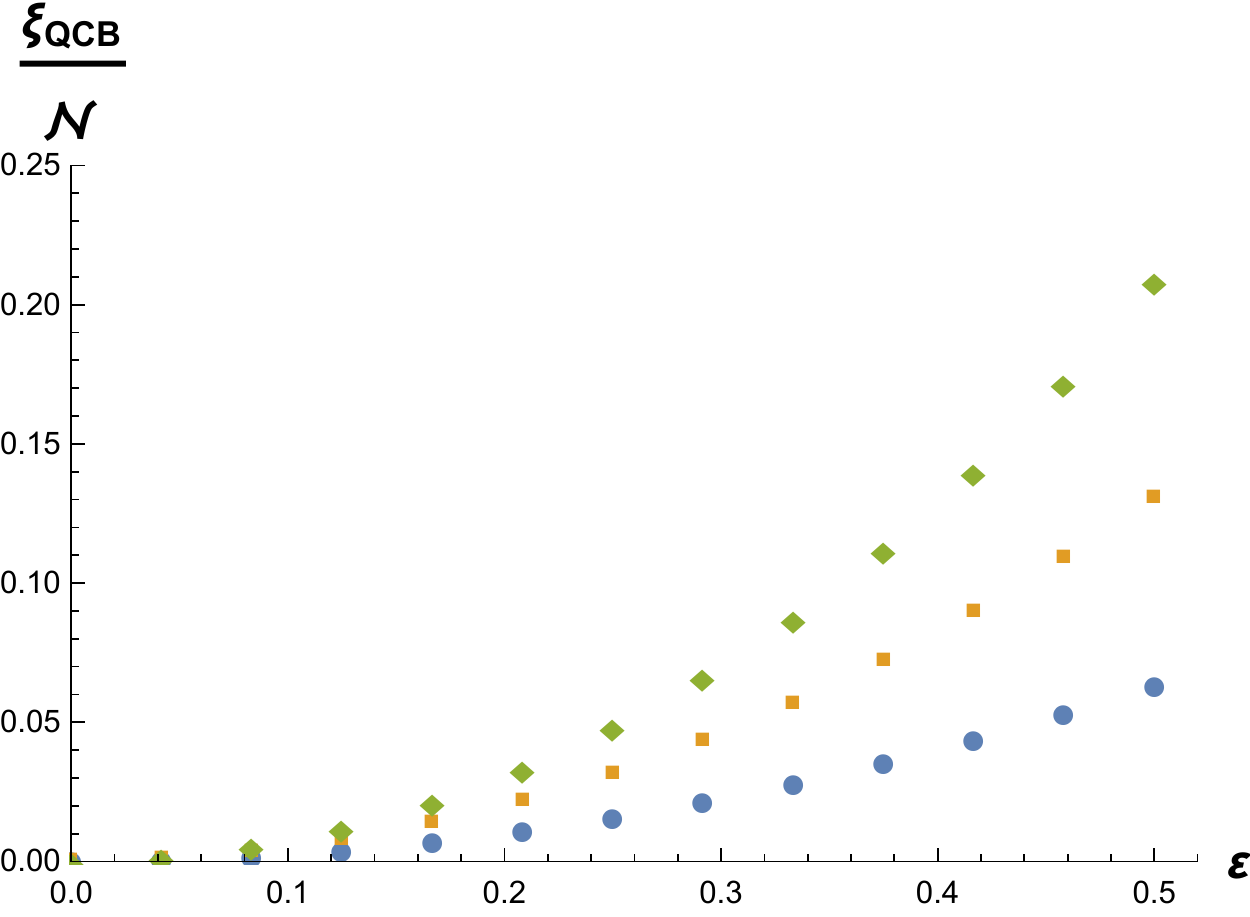}}
\caption{Quantum Chernoff bound for the phase gate $R_{{\bm e}_z}(\pi/10)$ and initial state $\ket{+}$ as a function of the deterministic error $ \varepsilon $ for the channel applied once (blue dots, with $N=30$ samplings of the output), twice (orange squares, with $N=15$ samplings of the output) or three times (green rhombi, with $N=10$ samplings of the output) before measurement. We plot $\xi_{\qcb}$ divided by $\Nit$ to take into account the reduction of the number of samplings if the total test time is fixed and the time for applying the channel dominates over the measurement time.}
\label{pi10rot}
\end{figure}

\subsection{Phase gate}\label{phasegate}
As an example, let us consider the phase gate $R_{{\bm e}_z}(\theta)$ discussed in Sec.~\ref{motiv}. The phase implemented experimentally may be faulty and differ from the nominal one by some amount $\varepsilon$. Our aim is to evaluate the quantum Chernoff bound $\xi_{\qcb}$ for the distinguishability of the output $\tau=\Phi_\text{ideal}(\rhoin)$ of the ideal channel and $\rho=\Phi_{\textrm{faulty}}(\rho_{\textrm{in}})$ of the faulty channel, with $\rhoin$ the input state. 
After optimizing over all possible pure input states, the quantum Chernoff bound \eqref{chernoff} between the output states from the ideal channel and the faulty one gives an upper bound to the distinguishability of the two channels.

We shall study both the case where the error $\varepsilon$ on the rotation angle is unknown but fixed, and the case where it has some probability distribution $P(\varepsilon)$; in the latter situation the output of the faulty channel is a mixed state given by $\int P(\varepsilon)\Phi_{\theta+\varepsilon}(\rho_{\text{in}})\mathrm{d}\varepsilon$.

\subsubsection{Deterministic phase error}\label{sec.fixer}
As noted in Sec.~\ref{motiv}, if one performs measurements in the adapted basis  $\{R_{{\bm e}_z}(\Nit\theta)\ket{+},R_{{\bm e}_z}(\Nit\theta)\ket{-}\}$, the faulty channel for small $\varepsilon$ leads to a probability for the outcome ``$-$'' given by $\simeq (\Nit\varepsilon/2)^2$, independent of $\theta$, implying an advantage $\propto \Nit$ for fixed total test time. Our numerical results for the quantum Chernoff bound corroborate this: In Fig.~\ref{pi10rot} we plot the quantum Chernoff bound \eqref{chernoff} for a small rotation angle $\theta=\pi/10 $ as function of the fixed unknown error $\varepsilon$. Clearly, for this value of $\theta$, the coherent enhancement of the distinguishability by iterating the channel occurs for all deterministic errors considered. 

Note that in the above reasoning we disregard the fact that the rotation $R_{{\bm e}_z}(\Nit\theta)$ of the measurement basis may suffer itself from errors, see \cite{nielsen_gate_2021}. 
The situation changes when considering a fixed measurement basis, say, the basis $\{\ket{+},\ket{-}\}$, irrespective of $\Nit$, see Appendix \ref{classchernoff}. 
The result for the classical Chernoff bound in this case is reported in Eq.~\eqref{defsmin}. Then, the optimal number of iterations depends on $\theta$, see Fig.~\ref{xy}. While for sufficiently small $\theta$ iterating the channel is always beneficial (within the three values $\Nit=1,2,3$ considered) as $\xi\sim\Nit^2$, this is no longer the case for larger values of $\theta$ (see discussion in the Appendix).

 \begin{figure}
\includegraphics[scale=0.6]{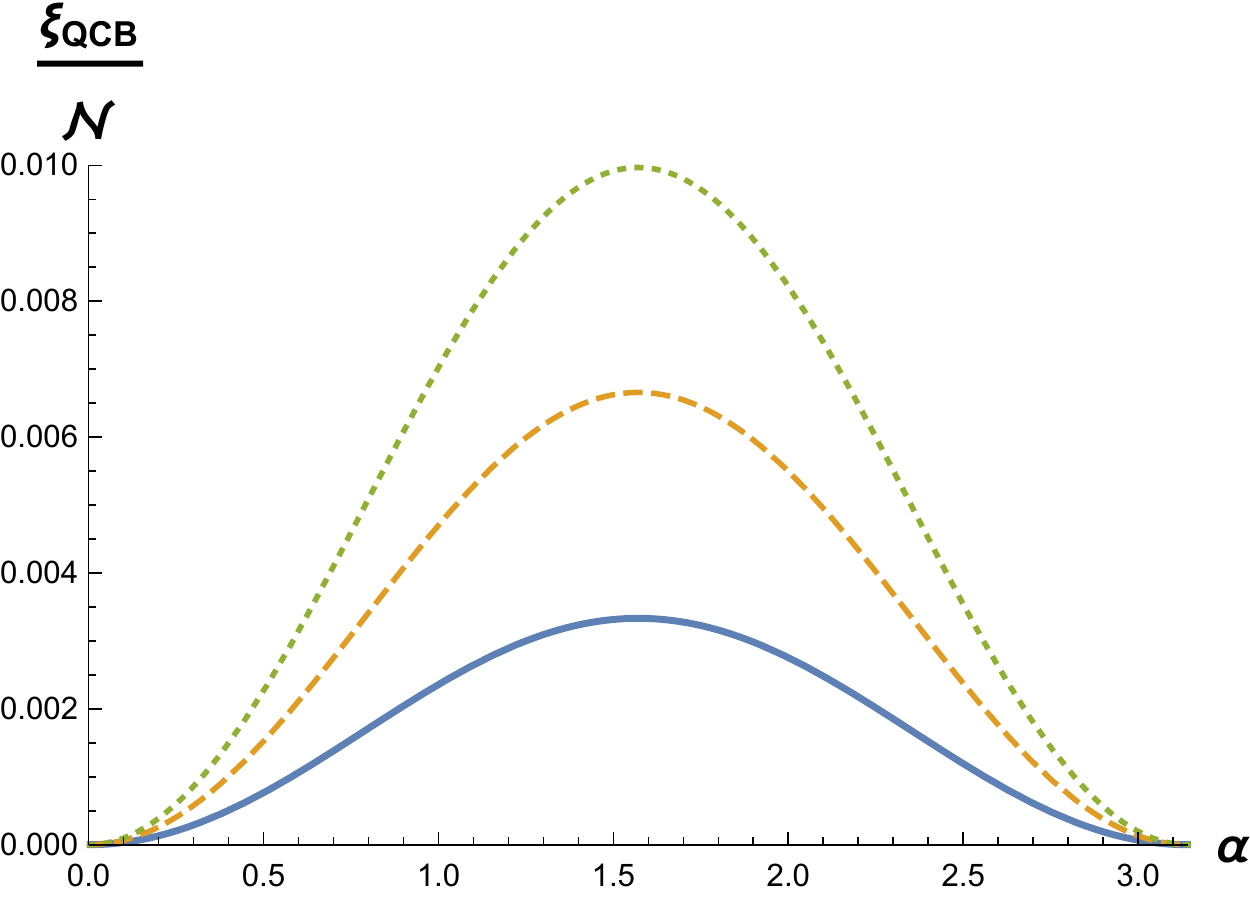}
\caption{Quantum Chernoff bound for the Z gate as a function of the angle $ \alpha $ parametrizing the input states. The error $ \varepsilon $ is uniformly distributed over the interval [0,1/5]. The blue curve corresponds to the gate applied once ($\Nit=1$), the orange one (dashed) to the gate applied twice ($\Nit=2$) and the green one (dotted) to the gate applied three times ($\Nit=3$) before each measurement. Same number $N_0=300$ of samplings of the channel, and $N=N_0/\Nit$ measurements of the output.}
\label{Zgate}
\end{figure}
\subsubsection{Random phase error}
\begin{figure}
\includegraphics[scale=0.6]{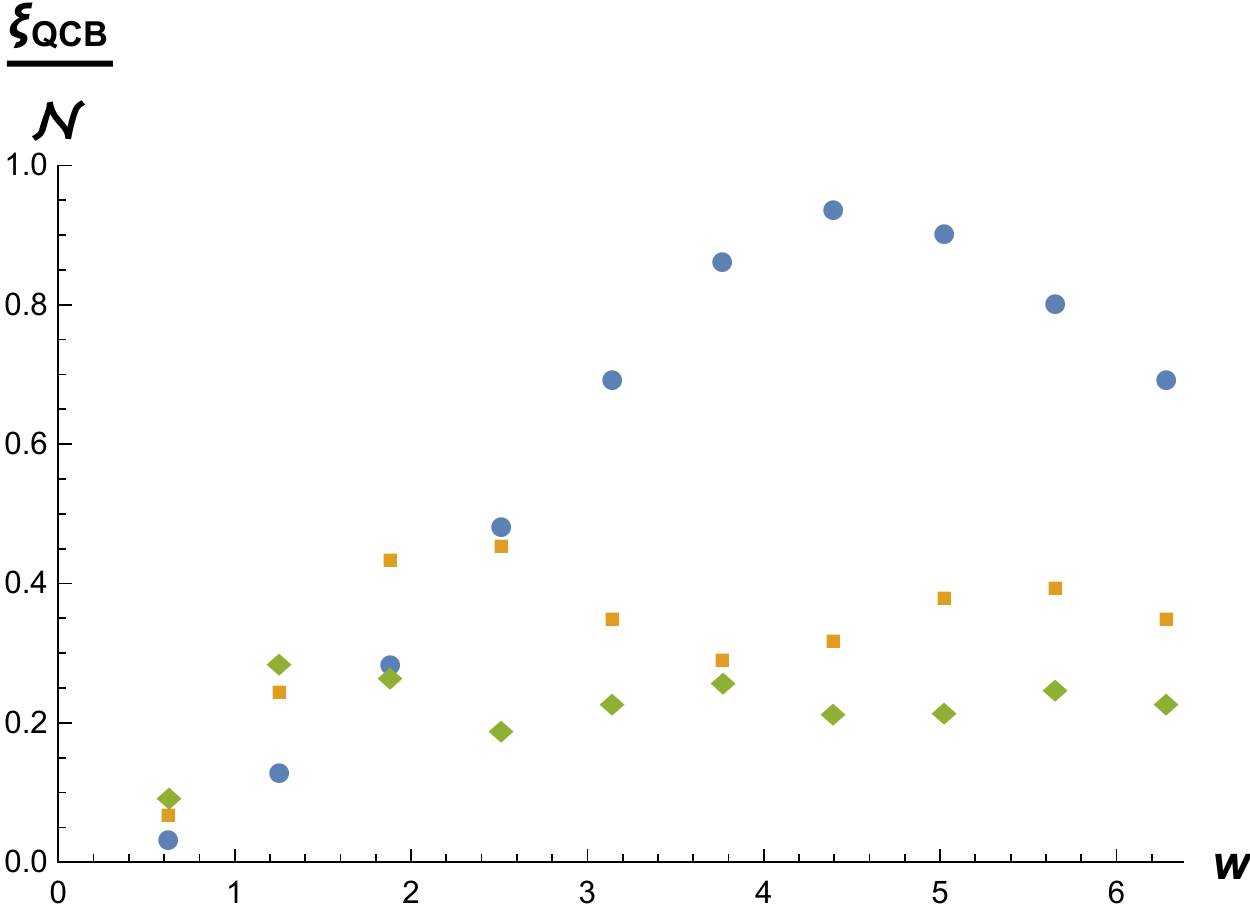}
\caption{Quantum Chernoff bound for the Z gate as a function of the width $w$ of a uniform distribution of errors. 
There are $N_0=30$ samplings of the channel, and $N=N_0/\Nit$ measurements of the output, with $\Nit=1$ (blue dots), $\Nit=2$ (orange squares), and $\Nit=3$ (green rhombi). 
}
\label{interval}
\end{figure}

Let us now consider a random phase error, uniformly distributed over an interval $[0,w]$, and suppose that the rotation angle is $\theta=\pi$ 
(thus the channel implements a $Z$ gate). In Fig.~\ref{Zgate} we compare the quantum Chernoff bound for $w=0.2$ and different numbers $\mathcal{N}$ of iterations of the channel, as a function of the angle $\alpha$ charaterizing the input state via its Bloch vector $\mathbf{r}_\textrm{in}$.
We find that for narrow distributions of errors the quantum Chernoff bound is always greater if the channel is iterated, up to three times, rather than applied only once, whereas for wide distributions, $w\gtrsim 2$, iterations can be disadvantageous, as shown in Fig.~\ref{interval}. Qualitatively, this might be understood by the fact that the coherent enhancement of the error is lost for such strongly mixed states, whereas the number of samplings is reduced by a factor $\Nit$ compared to the non-iterated case. 



\subsection{Phase gate with dephasing}\label{pgd}
The phase gate of Sec.~\ref{phasegate} can be realized by having a qubit evolve for a controllable time $t$ under the Hamiltonian $H(\omega)=\frac{\omega}{2}\sigma_{z} $ \cite{Granade2012, Ferrie2013,Fiderer2021}, giving $\theta=\omega t$ for $\hbar=1$. If in addition the qubit suffers from a dephasing
process with characteristic time $T_2$, the coherences decay exponentially as $\exp(-\gamma t)$ with $\gamma=T_2^{-1}$. We now assume that the error $\varepsilon$ affects the value $\gamma$ at which the true device operates. Just as for $\theta$, one would like to find as quickly as possible whether the true value $\gammat$ of $\gamma$ coincides with the value $\gamma_\textrm{c}$ within a given interval. Due to the decoherence process, it is intuitively less clear
whether iterating the channel helps. In order to assess whether iterating helps, we derive an exact expression for the quantum Chernoff bound. After time $t$, an initial pure state parametrized by angles $\alpha$ and $\beta$ becomes
\begin{equation}
\label{rhodeph}
    \rho_t(\gamma)=\begin{pmatrix}
      \cos ^2\frac{\alpha }{2} & \frac{\sin \alpha}{2}   e^{-\gamma t- i (\omega t+\beta )} \\
 \frac{\sin \alpha}{2} e^{-\gamma t+i (\omega t+\beta )} & \sin ^2\frac{\alpha }{2}
    \end{pmatrix}\,.
\end{equation}
If we want to distinguish channels with parameters $\gamma$ and $\gamma'$, the quantum Chernoff bound is  obtained from Eq.~\eqref{chernoff} by evaluating the infimum of $T=\tr[\rho_t(\gamma)^{s}\rho_t(\gamma')^{1-s}]$ over $s$. We define the dimensionless parameters $\tau=\gamma t$, $\varepsilon=(\gamma'-\gamma)/\gamma$,  $\tau'=\gamma' t=\tau(1+\varepsilon)$ and $\phi=\omega t + \beta$. It turns out that $T$ does not depend on the complex phase $\phi$, which can be therefore discarded.  Moreover $T$ is symmetric about $\alpha=\pi/2$, $T(\alpha)=T(\pi-\alpha)$. The value $\alpha=\pi/2$ is thus a local extremum. Numerically, the minimum over $s$ and $\alpha$ is always achieved at $\alpha=\pi/2$, which corresponds to a qubit initially prepared in the pure state $\ket{+}=\frac{1}{\sqrt{2}}(\ket{0}+\ket{1})$. If we fix $\alpha=\pi/2$ for the initial state, the output of the channel reads
\begin{equation}
\label{rhodetau}
    \rho(\tau)=\frac12\begin{pmatrix}
     1 &    e^{-\tau} \\
   e^{-\tau}&  1
    \end{pmatrix}\,.
\end{equation}
We now wish to distinguish channels with parameters $\tau$ and $\tau'$. Iterating the dephasing channel $\Nit$ times amounts to multiply $t$, and thus $\tau$ and $\tau'$, by a factor $\Nit$.

\begin{figure}
\includegraphics[scale=0.9]{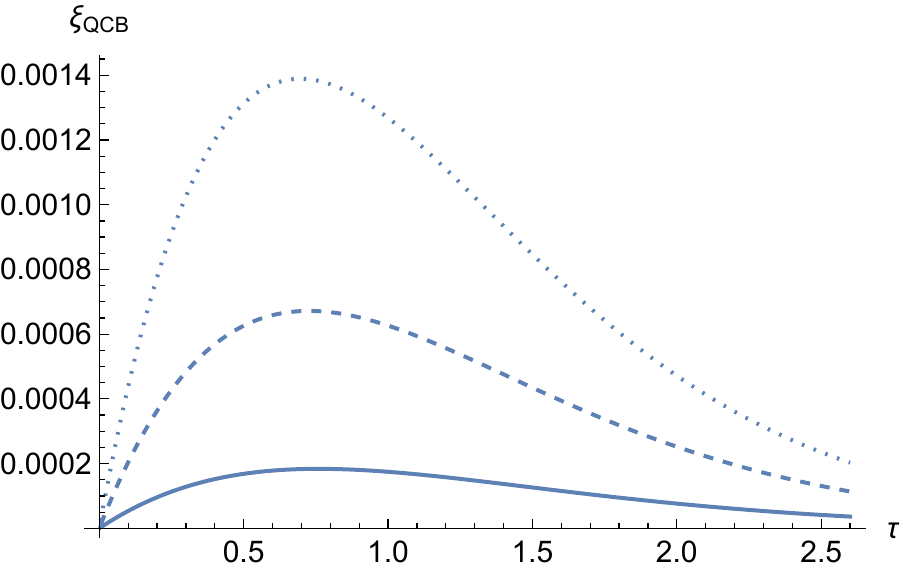}
\caption{Quantum Chernoff bound for the dephasing channel as
  function of $\tau$ for  (from bottom to top) $\varepsilon=0.1$ (solid), 0.2 (dashed), 0.3 (dotted). Note that varying $\Nit$ now amounts to varying $\tau$.
\label{xydephasing}}
\end{figure}

At fixed $\tau$ the quantum Chernoff bound is given by the infimum of $T(s)=\tr[\rho(\tau)^{s}\rho(\tau')^{1-s}]$ over $s$. Setting 
\begin{equation}
x=\frac{1+e^{-\tau(1+\varepsilon)}}{2},\qquad y=\frac{1+e^{-\tau}}{2}, 
\label{defxy}
\end{equation}
$T$ simplifies to
\begin{equation}
\label{Tpisur2}
  T=x^{1-s}y^s+(1-x)^{1-s}(1-y)^s\,.
\end{equation}
Remarkably, this expression exactly coincides with Eq.~\eqref{deff} giving the classical Chernoff bound in the context of a faulty rotation angle; only the definition of $x$, $y$ differs. The minimum of \eqref{Tpisur2} over $s$ thus yields a value $s_{\min}$ given by Eq.~\eqref{defsmin}, which provides an explicit analytic expression $\xi_{\qcb}=-\log T(s_{\min})$ for the quantum Chernoff bound of the dephasing channel. This quantity is plotted in Fig.~\ref{xydephasing}: as can be expected intuitively, the smaller the error $\varepsilon$, the larger the number of measurements needs to be to distinguish the faulty from the ideal channel.

We can now address the question whether iterating the channel helps or not. For $\Nit$ iterations of the channel $\tau$ is changed to $\Nit \tau$, and a series expansion of $T(s_{\min})$ at small ${\varepsilon}$ gives 
\begin{equation}
 \label{xipetiteps}
\frac{\xi_{\qcb}}{\Nit}=\frac{\Nit\tau ^2 {\varepsilon} ^2}{8 \left(e^{2 \Nit\tau }-1\right)}+O\left({\varepsilon} ^3\right)
\end{equation}
for the quantum Chernoff bound in units of the number of iterations. At fixed $\varepsilon$ and $\tau$, \eqref{xipetiteps} is a decreasing function of $\Nit$.
This means that iterating the channel leads to a slower decrease of the error probability with the number of steps. This is in contrast with the case of the rotation channel of the previous section, where the QCB is systematically smaller for smaller $\Nit$ (see Fig.~\ref{Zgate}), and thus iterating the channel increases the success probability. A way of understanding this intuitively is to observe that following Eq.~\eqref{rhodetau}, $\rho(\tau)$ goes to the maximally mixed state as $\tau$ increases. Therefore, at large values of $\tau$ (which corresponds to larger $\Nit$) states $\rho(\tau)$ and $\rho(\tau+\varepsilon)$ both flow towards the same state and become less and less distinguishable.

\section{Bayesian experimental design} \label{ngu}

The previous section was based on the frequentist approach to parameter estimation.  However, in the light of wanting to speed up the decision problem, it is natural to use Bayesian parameter estimation combined with optimal experimental design strategies: Instead of waiting for measurement results from a possibly large
number of measurements, each measurement can be used to update the knowledge about the unknown parameter. Any past measurement outcome can be used for determining which subsequent measurement promises the largest information gain, given the current knowledge about the parameter. In Sec.~\ref{sec.it} we discussed the advantage that can be expected from iterating the quantum channel; in the Bayesian approach, the option to iterate the channel without measurement can easily be included by considering the identity operator as a possible POVM. It is then the information already gathered that determines at each step if it is more
advantageous to continue iterating the channel or to do a measurement. This naturally leads to ``non-greedy'' Bayesian updates and experimental designs, where one tries to find an optimal sequence of measurements rather than just the next one. We briefly review the Bayesian methods and adaptive strategies, adapt them to the present context of quantum functional testing, and then study their performance.

\subsection{Bayesian methods and adaptive strategies}
\label{smc}
At the core of the Bayesian methods of parameter estimation is the Bayes rule for conditional probabilities $\prob(\mathbf{x}|\mathbf{D})=\prob(\mathbf{D}|\mathbf{x})\prob(\mathbf{x})/\prob(\mathbf{D})$, where $ \mathbf{x} $ is the vector of unknown parameters and $\mathbf{D}=\{d_1,d_2,...,d_n\} $ are the data gathered from $n$ experiments $ \mathbf{e}=\{e_1,e_2,...,e_n\} $. In the above formula, $\prob(\mathbf{x}) $ is called the \emph{prior} and it incorporates any a priori knowledge on the parameters, $ \prob(\mathbf{x}|\mathbf{D})$ is referred to as the \emph{posterior}, $ \prob(\mathbf{D}|\mathbf{x}) $ is the \emph{likelihood}, which in the present context is given by the Born rule, and $\prob(\mathbf{D})$ is a normalization factor. \\ 
An analytic calculation of the posterior probability distribution soon becomes impractical after few experiments/updates. Thus numerical methods, often involving Monte Carlo algorithms, have been implemented to overcome these difficulties. We consider here the so-called \emph{sequential} Monte Carlo algorithms (SMC) \cite{Doucet2000}, where data are processed sequentially, that is, the posterior at step $k$ is taken as the prior at step $k+1$. 
In the "particle filtering" approach to SMC, the prior continuous probability distribution $ \prob(\mathbf{x}) $, corresponding to a case where parameter space is continuous, is replaced with a discrete approximation that can be written as $\sum_i w_i\delta(\mathbf{x}-\mathbf{x}_i)$, where the $\mathbf{x}_i$ are the locations of "particles" (which sample the space of parameters) and $ w_i $ the associated "weights" (probabilities). A particle filter $ \{w_i',\mathbf{x}_i'\} $ for the posterior probability distribution $\prob(\mathbf{x}|\mathbf{D})$ can be obtained by setting $\mathbf{x}_i'=\mathbf{x}_i$ and updating the weights from the Bayes rule as $w'_i=w_i \prob(\mathbf{D}|\mathbf{x_i})/\sum_j w_j\prob(\mathbf{D}|\mathbf{x_j})$. This update method suffers from numerical instability as data are collected, a problem known as impoverishment of the filter.   
Resampling 
new particle locations can solve this problem \cite{Doucet2009}. A resampling algorithm substitutes the impoverished filter with a new enriched one that still approximates the same probability distribution, only modifying the choice of particle locations $\mathbf{x}_i$.

Including new knowledge coming from the data through the update of the conditional posterior probability distribution suggests adaptive strategies to choose the next best experiment. A lot of effort has been put in Bayesian experimental design \cite{Chaloner1995}, which recasts the problem of choosing the best experiment into the theoretical problem of optimizing a \emph{utility function} $U$. At step $n$, the function $U$ depends on the possible future $m$ designs/experiments $ \mathbf{e'}=\{e_{n+1},e_{n+2},...,e_{n+m}\} $, the possible future data $\mathbf{D'}=\{d_{n+1},d_{n+2},...,d_{n+m}\} $ and the model parameters $ \mathbf{x} $. One may consider the optimal Bayesian design $ \mathbf{e'}^* $ among the ones in the design space $ \mathbf{E}$. If $U(\mathbf{e'},\mathbf{x},\mathbf{D'})$ is the utility function, then the optimal design $\mathbf{e'}^*$ is the one that maximises the expected utility 
\cite{Ryan2016, Granade2012},
\begin{equation}
\mathbf{e'}^*=\argmax_{\mathbf{e'}\in \mathbf{E}} E[U(\mathbf{e'},\mathbf{x},\mathbf{D'})]\,,
\label{utility}
\end{equation}
where $ E[...] $ denotes the expectation value of the utility function over the possible outcomes of the experiments.

\subsection{Greedy vs non-greedy designs}\label{greedy}
If the above optimization is performed at each step, this gives a \emph{sequential} or \emph{adaptive} design. Many adaptive experimental designs involve optimizing only over a single future experiment and outcome $\{e_{n+1},d_{n+1}\}$; this case where $m=1$ is called the \emph{myopic} (or \emph{greedy}) approach. However, an update can occur in the optimization after one or more actions, i.e.~$ m\geq 1 $. Even classically, the case $m=1$ is in general not optimal, since one could look ahead to \textit{all} future observations and decisions that can be made at each step in the experiment. For quantum functional testing, this $m\geq 2$ approach appears even more attractive due to the possibility of coherent enhancement of errors discussed above. Optimizing the experimental design over several steps is
known as \emph{backward induction} \cite{Bernardo2009}. It is computationally very expensive due to the exponential proliferation of combinations, and hence
remains limited to relatively simple design problems. Here we use an intermediate strategy, which goes beyond the single step optimization, but is limited to a modest  number of steps ahead ($ 1\leq m\leq 4 $).

Common choices of utility functions $U$ for parameter estimation problems are functions of the posterior covariance matrix or information-based utilities
\cite{Ryan2016, Granade2012}. Going from prior to posterior distribution should increase the available information. The information gain ($\ig$) \cite{Lindley1956} is defined via the Kullback-Leibler divergence (KLD) between the prior distribution $\prob(\mathbf{x})$ (omitting the dependence on earlier experiments $ \mathbf{e} $ and outcomes $ \mathbf{D} $) and the posterior distribution $\prob(\mathbf{x}|\mathbf{e'},\mathbf{D'})$ where we gained some knowledge $ \mathbf{D'} $ from the experiments $ \mathbf{e'}$. The information gain resulting from these additional experiments is
\begin{align}
\begin{aligned}
\ig(\mathbf{e'},\mathbf{D'})
&=\int_{\mathbf{X}}\prob(\mathbf{x}|\mathbf{e'},\mathbf{D'})\log\frac{ \prob(\mathbf{x}|\mathbf{e'},\mathbf{D'})}{\prob(\mathbf{x})} d\mathbf{x}\\
&\equiv E_{\mathbf{x}|\mathbf{e'},\mathbf{D'}}
\left[\log\frac{ \prob(\mathbf{x}|\mathbf{e'},\mathbf{D'})}{\prob(\mathbf{x})}\right]\,.
\label{kld}
\end{aligned}
\end{align}
In \eqref{kld} we take the integral over the sample space $\mathbf{X}$ of possible model parameters $\mathbf{x}$, and the index of $ E $ specifies the probability distribution to be integrated over for the calculation of the expectation value: in the present case, $E_{\mathbf{x}|\mathbf{e'},\mathbf{D'}}$ implies to use the probabilities $ \prob(\mathbf{x}|\mathbf{e'},\mathbf{D'}) $. 

We now define the mutual information (MI) as the information gain averaged over all possible outcomes \cite{cover1999elements}. Suppose each of the $m$ experiments has a finite number $ l $ of possible outcomes. The MI at fixed $m$ is then given by taking the integral over the data $\mathbf{D'}$, conditional on the design $ \mathbf{e'} $, hence by the expected value 
\begin{equation}
\textrm{MI}(\mathbf{e'})\equiv \overline{\ig}^{(m)}=\sum_k p_k \ig(\mathbf{e'},\mathbf{D'}_k)
\label{avIG}
\end{equation}
where $ k $ runs up to the total number $l^m$ of possible outcome sequences, $p_k $ is the probability of obtaining the $k$th sequence of outcomes, and $\ig(\mathbf{e'},\mathbf{D'}_k)$ is the information gain of the $k$th sequence. For a channel $\Phi_\mathbf{x}$ depending on parameter $\mathbf{x}$ the probability $p_k$ is given by the Born rule for the state $\Phi_\mathbf{x}(\rho_{\textrm{in}})$. Namely, if experiments $\mathbf{e'}=\{e_{n+i},1\leq i\leq m\}$ are realized by POVMs with operators $M_{i,j}$ (with outcomes labeled by $j$ for experiment $e_{n+i}$), then the $k$th sequence of outcomes $\mathbf{D'}_k=\{j_1,j_2,...,j_m\}$ is obtained with probability
\begin{equation}
  \prob_{\mathbf{e'}}(\mathbf{D'}_k|\mathbf{x})=\tr (T_k^\dagger T_k \Phi_\mathbf{x}(\rho_{\textrm{in}})),\quad
  T_k=M_{m,j_m}...M_{2,j_2}M_{1,j_1}
\end{equation}
\cite{Nielsen00}, 
and the probability $p_k$ then reads
\begin{equation}
p_k=\prob_{\mathbf{e'}}(\mathbf{D'}_k)=\int_{\mathbf{X}}\prob_{\mathbf{e'}}(\mathbf{D'}_k|\mathbf{x})\prob(\mathbf{x})d\mathbf{x}.
\end{equation}
Note that in this case the prior distribution $\prob(\mathbf{x})$ in Eq.~\eqref{kld} is not the one relative to the previous single step, but rather the one of $m$ steps back.

In Sec.~\ref{applications} we shall compare the efficiency of two different utility functions $U$, the negative variance of the posterior distribution (VAR) and the mutual information (MI), and see how they perform in the case of single parameter estimation.

\subsection{Experimental design and decision criteria}\label{design}
Our general framework is the following. We conceive a scenario in which a company provides us with a device which realizes a quantum channel depending on some parameter  $\mathbf{x}$. The company claims that the nominal value of $\mathbf{x}$ is within the \emph{confidence region} $\mathbf{x}_\textrm{c}\pm\Delta\mathbf{x}$, where $ \Delta\mathbf{x} $ defines a small interval around each component of the vector $ \mathbf{x}_\textrm{c} $. 
Our aim is to establish as quickly as possible whether the vector of true parameters $\xt$ characterizing the channel lies in that confidence region. In the case of a single parameter $ x_\textrm{c} $ this region is simply an interval centered around $ x_\textrm{c} $; in the following we refer to this interval as ``spec'' (for ''specification by the company''). The functional test process consists of two stages; in the first one, we update the probability distribution of $\mathbf{x}$ following an adaptive strategy, and in the second one we use a \emph{decision criterion} to accept or reject the spec given by the company, and reach a conclusion.\\

We consider two different decision criteria. The first criterion (which we label MEAN) only relies on the position of the mean of the posterior distribution obtained, looking whether it falls inside or outside the spec. The second criterion compares the spec interval with a chosen \emph{credible region} \cite{Granade2012,Ferrie2014,Bernardo2005} for the posterior distribution (intuitively, the region where most of the probability lies). We took the \emph{highest posterior density} (HPD) region as the credible region, which in our case is defined as the region containing $ 95\% $ of the probability mass;  the "HPD" criterion is defined by accepting the value $ x_\textrm{c} $ provided by the company if $ 95\% $ of the HPD credible region falls inside the spec, and rejecting it if  $ 95\% $ falls outside. 
In the remaining cases this criterion remains inconclusive, and we count these instances as failed tests, i.e.~they contribute to the estimated error probability. The overlap of this credible interval with the spec is central to the decision method used in \cite{Kruschke2018}.

\begin{figure*}
\includegraphics[scale=0.6]{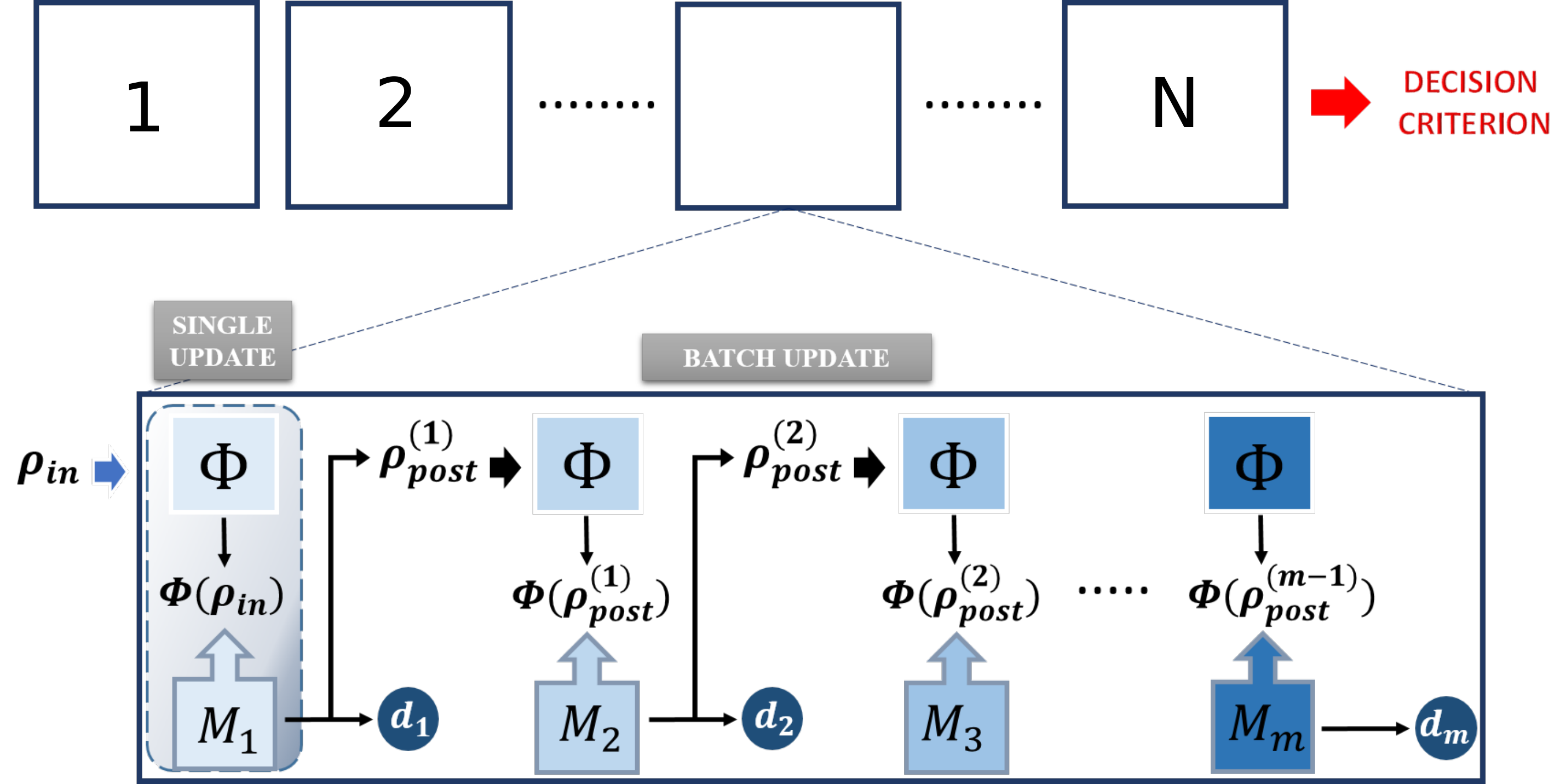}
\caption{Sketch of the certification protocol in the
  experiment-limited scenario. In a batch update the state between
  single updates is not reset to the initial input state. The obtained
data $d_1,\ldots,d_m$ are used to find the next set of $m$
measurements (including possibly the identity) that maximizes the
expected increase of information.} 
\label{batch}
\end{figure*}

\section{Application to quantum channels}\label{applications}

We now apply this scheme to the problem of certifying quantum channels governed by a single parameter, focusing again on the phase gate and the dephasing channel. The best sequence of measurements of length $ m $ is the one which maximises the utility function $U$. We denote with $\mathcal{M}$ the set of measurements one can choose from; each measurement is performed on the output state given by the application of the channel on an input state. Besides the measurements in $\mathcal{M}$, we introduce the possibility of choosing an extra action, which consists in applying again the channel over the output state instead of measuring; this is implemented in practice by including the identity operator in the set $\mathcal{M}$. The choice of this action entails that the input state is not always reset to the initial one between single actions (measurement or identity channel). We assume that the certification process is experiment-limited, which means that the number of actions $N_0$ is fixed, while the number of updates of the posterior distribution depends on $ m $ as $N=N_0/m$. We sketch this certification protocol in Fig.~\ref{batch}.

In what follows, we analyze and compare the choice of $U$ (either MI or VAR) and the role of the decision criterion (MEAN or HPD).

\subsection{Phase gate}

\subsubsection{Setup}
\label{1stmodel}
Let us consider the phase gate $R_{{\bm e}_z}(\theta)$ defined in Sec.~\ref{sec.it}. We fix the initial input
state as $\rho_{\textrm{in}}=\vert + \rangle \langle + \vert$, whose Bloch vector points along the $ \hat{x} $ axis. 
We consider an experiment which involves two-outcome measurements, given by two POVM elements, e.g.~projectors $\Pi_1^{(i)}=(\mathbb{I}+\sigma_i)/2$ (and $\Pi_2^{(i)}=\mathbb{I}-\Pi_1^{(i)}$), with $\sigma_i$ the Pauli matrices $ \sigma_x $ and $\sigma_y$. 
The probability of obtaining one of the two outcomes is given by the Born rule $ p_{1,2}^{(i)}=\tr(\Pi_{1,2}^{(i)}\rho_{\textrm{out}}) $, where at the first update $\rho_{\textrm{out}}=\Phi(\rho_{\textrm{in}})$. 

To simulate an experiment, we choose as initial prior a uniform distribution of $\theta$ over the $[-\pi,\pi]$ interval.  We fix the true angle $\thetat$, and the angle $\theta_\textrm{c}$ and a spec around it provided by the company as $[\theta_\textrm{c}-\Delta\theta_\textrm{c},\theta_\textrm{c}+\Delta\theta_\textrm{c}]$. 
According to the central limit theorem (or to the Bernstein-von Mises theorem in the context of Bayesian inference) the posterior distribution
tends towards a normal distribution for $ N\gg 1 $ measurements, and the standard deviation  of the posterior distribution decreases as $
\sigma=\sigma_{\rm prior}/\sqrt{N} $ for large $N$. 
A possible choice for the number of measurements after which one can decide to stop the update process and apply the decision criterion is that $\sigma$ becomes comparable with the standard deviation $\sigma_{\rm spec}$ of a uniform distribution inside the spec. This corresponds to $ N_0=\sigma^2_{\rm prior}/\sigma^2_{\rm spec}$ measurements for $m=1$.

The set $\mathcal{M}$ one can choose from at each step includes the POVM elements defined above, as well as the identity matrix, corresponding to simply iterating the channel without measurement. To compare the efficiency of the different optimization methods (corresponding to different choices of $m$  and utility function $U$), we
calculate the probability of success, that is, the probability of reaching the correct conclusion, for the two  decision criteria discussed in Sec.~\ref{design}. 
 In particular, since we fixed the total number of actions, we also consider as not successful for the "HPD" criterion the cases in which it remains inconclusive.

\subsubsection{Numerical results}
We consider sequences of measurements of length $m=1$ to 4, and both utility functions -- variance of posterior and mutual information -- discussed above. Our computations are performed with the Qinfer package \cite{Granade2017} which uses the SMC methods discussed in Section \ref{smc}. In this package, the credible region used in the HPD criterion is defined by ordering particles by decreasing weight and selecting the region including the highest-weight particles such that the desired credibility level ($ 95\% $) is reached. This corresponds to the HDP region defined above. In principle, there could be an issue with the connexity of the support (for instance if the distribution is bimodal). This has been discussed e.g.~in \cite{Ferrie2014b}. Here, in practice, posterior distributions always end up having Gaussian-like shapes, and thus the above construction of the credible region meets its purpose. We used a particle filter with 2000 particles, and we checked that for this choice the numerical values of the MI agree up to the second decimal digit with the analytical result obtained after one Bayesian update of the probability distribution.

\begin{figure}
\includegraphics[scale=0.45]{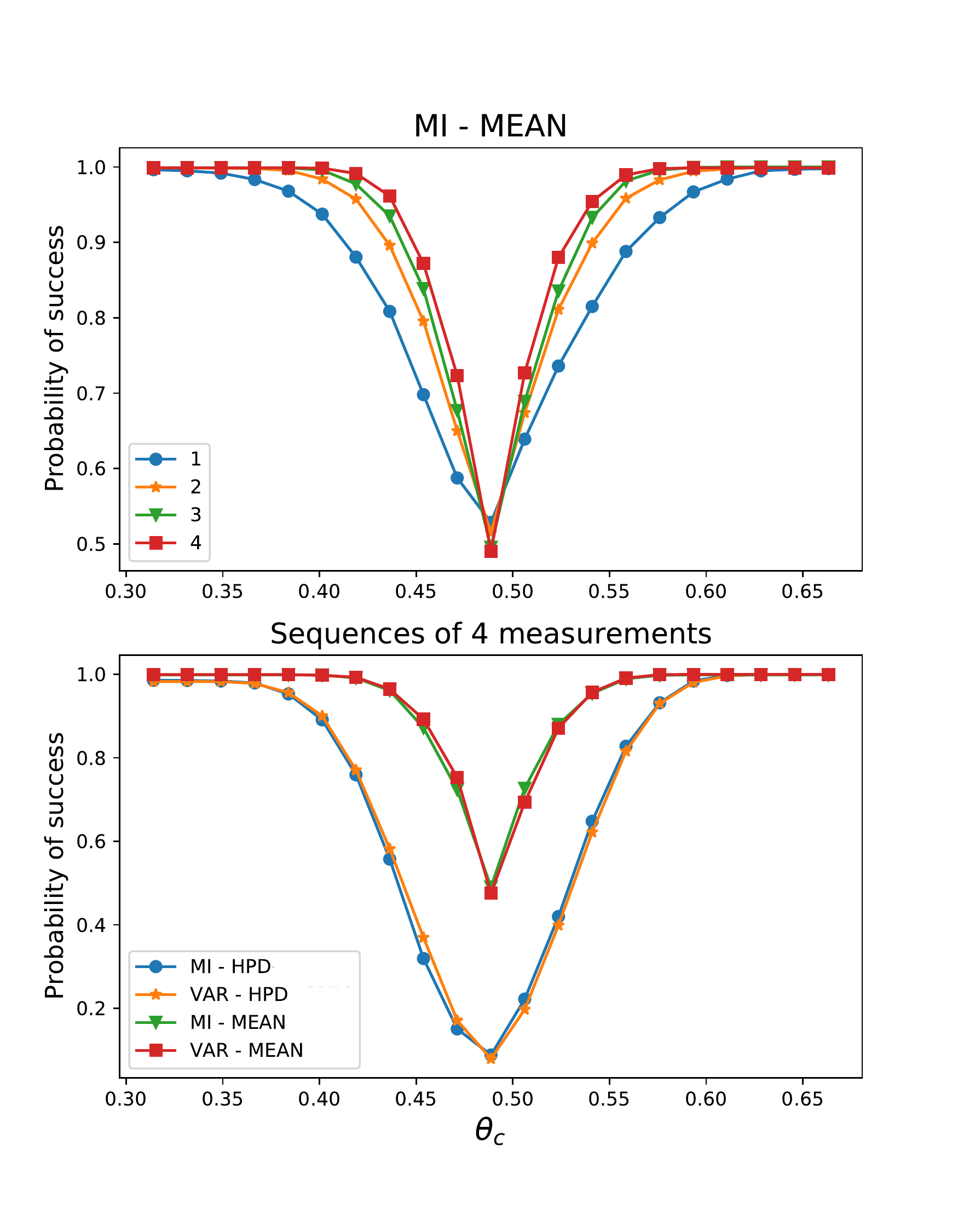}
\caption{(Upper panel) Success probabilities for correctly classifying the phase gate  $R_{{\bm e}_z}(\theta)$ as in or out of spec concerning $\theta$
for different lengths of measurement sequences. Here we consider $\thetat=\pi/10$ and $\Delta\theta_\textrm{c}=\pi/18$. Utility function: mutual information (MI); decision criterion: position of the estimated mean value (MEAN). 
Probabilities are averaged over 2000 random particle filters with 2000 particles each.
The probabilities increase going from length 1 to 4. The minimum is reached when $ \thetat \simeq\theta_\textrm{c}-\Delta\theta_\textrm{c} $. Statistical fluctuations of the estimates of $p$ as judged from the minimum, which should be at $p=0.5$, amount to $ \sim 0.05 $. (Lower panel) Success probabilities for measurement sequences of length 4 for all four combinations of the two utility functions MI and VAR, and decision criteria HPD and MEAN. 
}
\label{newpanel}
\end{figure}

In Fig.~\ref{newpanel} we show the results obtained for a fixed angle $ \thetat=\pi/10 $ and several values of $ \theta_\textrm{c} $ with a fixed $\Delta\theta_\textrm{c}= \frac{\pi}{18}$. In order to reach a width of the posterior distribution of the order of the width of the spec we take $ N_0\approx 300 $. We find that the optimization over sequences of measurements of length $m\geq 2$ gives an advantage over the greedy approach $m=1$, both for the negative variance and for the mutual information. Namely, the final posterior distribution for $m\geq 2$ is narrower than the one obtained for $m=1$ with the same number of actions (measurements or iterations of the channel), despite a smaller number of Bayesian updates. After a few initial oscillations, the best sequence of actions always coincides with the one which contains $m-1$ times the identity and a single measurement ($\sigma_x$ or $\sigma_y$). By contrast, in the greedy approach, the identity measurement is never chosen, as it does not lead to any information gain by itself. \\

In the upper panel of Fig.~\ref{newpanel}
we present the results for the MI as utility function and MEAN as decision method, comparing sequences of $m=1,2,3 $ and $ 4 $ measurements.
The probability of success decreases and reaches the minimum value when the true angle $ \thetat$ almost coincides with one of the boundaries of the spec region (i.e.~when $ \thetat\simeq \theta_\textrm{c}\pm\Delta\theta_\textrm{c} $). In that case the posterior distribution is expected to converge to a Gaussian centered on the boundary; the mean fluctuates around the boundary from one random realization to the other, yielding only a $50\%$ chance of reaching the correct conclusion from the results of the decision criterion based on the position of the mean. The error made by having only a limited number of realizations can be estimated by considering the fluctuations of the mean at that point, where the theoretical value of the success probability is exactly $0.5$.  The difference between the largest and smallest values for different $\Nit$ is $ \sim 0.05 $, yielding an indication of how large statistical errors bars are for this point. We can see that the probability of success grows going from sequences of measurements of length 1 (greedy approach) to 4.  

In the bottom panel of Fig.~\ref{newpanel} we compare the results obtained for sequences of length $m=4$ for all four combinations of the utility functions MI, VAR and final decision criteria MEAN, HPD. We see that MEAN always gives greater success probabilities than the HPD criterion, whereas the utility functions has little influence. We expect bad performance of the HPD criterion when $\thetat$ is close to $\theta_\textrm{c}\pm \Delta_\textrm{c}$ as the final distribution has insufficient overlap with both the spec and the interval outside the spec, i.e.~the criterion remains inconclusive and the test is counted as unsuccessful. This is confirmed in Fig.~\ref{newpanel} for $\thetat=\theta_\textrm{c}- \Delta_\textrm{c}$. On the other hand, the MEAN criteria leads to essentially random classification at the border of the spec, resulting in about 50\% success rate (but also 50\% of wrongly classified gates). Depending on the use case either strategy can be more advantageous, the first one being safer, the second leading to less waste of devices still in spec. \\

\subsection{Phase gate with dephasing}

\subsubsection{Setup}
We now consider the phase gate with dephasing, defined in Sec.~\ref{pgd}. Suppose that we want to estimate the decay parameter $\gamma$. The controllable quantities are the evolution time $t$, which we fix to some constant value (or to a multiple of it when the channel is iterated), and the measurements to perform after the evolution, which are chosen with an adaptive strategy as in Sec.~\ref{1stmodel}.

We again choose the experiment-limited scenario, where $N$ is fixed, and we investigate whether iterating the channel, represented by the time evolution, can bring an advantage in drawing the right conclusion about the parameter to estimate.

\subsubsection{Numerical results}
\begin{figure}
\includegraphics[scale=0.45]{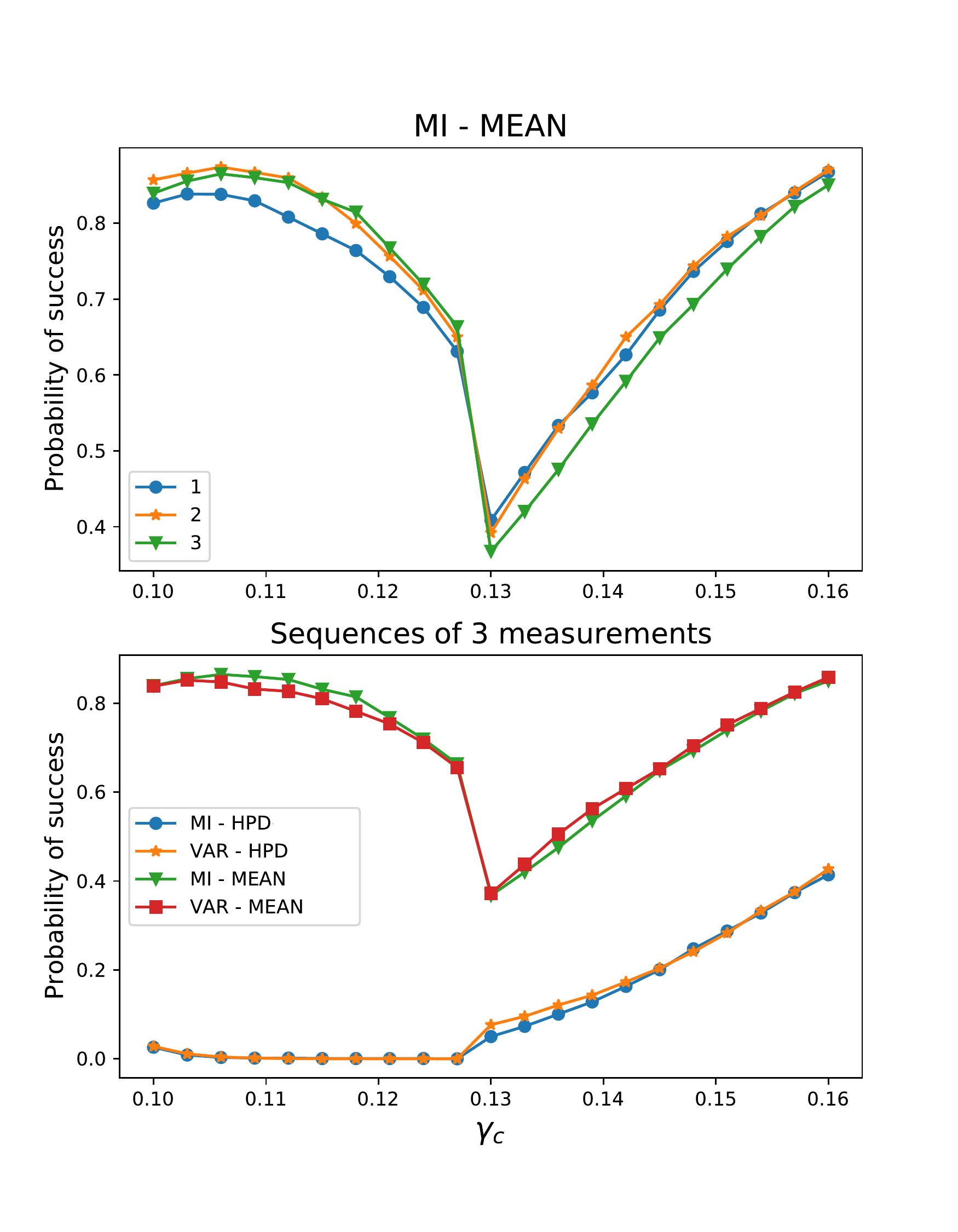}
\caption{(Upper panel) Success probabilities for correctly classifying the phase gate with dephasing as in or out of spec concerning $\gamma$ for different lengths of measurement sequences. Here we consider $\gammat=0.1$ and $\Delta\gamma_\textrm{c}=0.03$, so that for $\gamma_\textrm{c}<0.13$ $\gammat$ is within the spec and outside for $\gamma_\textrm{c}>0.13$. Utility function: mutual information (MI); decision criterion: position of the estimated mean value (MEAN). The probabilities are averaged over 2000 random particle filters with 2000 particles each and $N_0=300$.
The success probabilities improve only slightly from $m=1$ to $m=2$ and for $\gamma_\textrm{c}\lesssim \gammat+\Delta\gamma_\textrm{c}$. Then they coincide, showing no advantage in iterating the channel. Statistical fluctuations of the estimates of $p$ as judged from the minimum, which should be at $p=0.5$, amount to $ \sim 0.05 $.  (Lower panel) Success probabilities for measurement sequences of length 3 for all combinations of the utility functions (MI and VAR) and decision criteria (HPD and MEAN) considered. 
}
\label{panel2}
\end{figure}

To run the certification protocol we chose a uniform prior distribution of $\gamma$ over the interval $[0,1]$. We allow for the same two-outcome measurement as in Sec.~\ref{1stmodel}, and fix again the number of actions to $N_0=300$. 
We set a spec $\gamma\in[\gamma_\textrm{c}-\Delta\gamma_\textrm{c},\gamma_\textrm{c}+\Delta\gamma_\textrm{c}]$ with $\Delta\gamma_\textrm{c}=0.03$ provided by the company. We simulate an experiment with $\gammat=0.1$ and $t=5$ (for $k$ iterations of the channel we take $t=5k$). The adaptive strategy based on the two different utility functions never selects as optimal sequences of measurements the ones containing the identity, so never allowing for more than one application of the channel. We show the results obtained for the probability of success in Fig.~\ref{panel2}. 
In the upper panel we show that the probability of success is larger for $m=2$ or $3$ for the mutual information strategy with the decision criterion based on the mean as long as the true value $\gamma_t$ is inside the spec, whereas for $\gamma_t$ outside the spec, $m=3$ turns out to perform worse than $m=1,2$. This is explained by the fact that the posterior probability distributions have not completely converged to their asymptotic form for $N_0=300$. We show in the Appendix that for $N_0=1200$ the order is as expected, with $m=2,3$ performing better than $m=1$. 
In the bottom panel we show the difference between the probabilities of success for sequences of length 3, finding as in the case of the rotation angle (see Fig.~\ref{newpanel}) that both the utility functions perform better with the 'MEAN' as final decision criterion.
We showed in Fig.~\ref{xydephasing} that iterating the dephasing channel does not enhance the distinguishability between the ideal and faulty channels. Indeed, for $\gamma=0.1$ and $t=5$, the corresponding value $\tau$ is in a regime in Fig.~\ref{xydephasing} where $\xi_{\rm QCB}$ is not a decreasing function of $\tau$.
Fig.~\ref{panel2n01200} for $N_0=1200$ shows that iterations slightly enhance the probability of success for the correct identification of the channel (ideal or faulty). This might be seen to contradict the findings based on the quantum Chernoff bound.  However, quantum Chernoff bound and success probabilities give answers to two distinct questions. If one wishes to distinguish channels, they should be as far apart as possible in the space of states, i.e.~$\Nit$ should be as small as possible according to Eq.~\eqref{rhodetau}. However if the question is to detect as fast as possible whether a probability distribution is within or outside some spec interval, it is their distance in the space of the parameter we are probing that matters: in that case, for a fixed number $N_0$, larger $\Nit$ may correspond to a faster decrease in the width of the posterior distribution, thus to a higher distinguishability.

\section{Conclusions}
To summarize, we have described how  the perspective of quantum functional testing leads to a new twist of quantum parameter estimation, namely one where the rapid gain of information about a parameter and a decision problem to accept or reject a device is center stage, rather than ultimate sensitivity. We have introduced several ingredients that allow one to speed up the decision problem, most importantly the possibility of coherent amplification of errors by iteration of a quantum channel before measurement. We have analysed the interest of iterations first with Chernoff bounds, both classical and quantum, and found that for single qubit gates and deterministic (or narrowly distributed) errors in the rotation angles with negligible decoherence, iterations of the channel tend to be beneficial, where however attention has to be paid to choosing appropriately adapted measurements as well. We then introduced non-greedy Bayesian updates and experimental design, where an agent decides on the fly whether it is more advantageous to continue iterating the quantum channel or measuring, and what kind of measurement to perform, based on the previously accumulated data. This setup was investigated for two different utility functions and two decision criteria.  The simulations showed that the estimate of rotation angles profits from iterating the quantum channel.  In contrast, for the dephasing parameter the situation is more complex and depends on whether or not the true value of the parameter is inside or outside the spec.  We also stressed the fact that assessing whether the parameter value is within a certain interval or not is distinct from the problem of distinguishing two quantum channels. 

\section*{Acknowledgments}
This work is funded by the Ministry of Economic Affairs, Labour and Tourism Baden W\"urttemberg in the frame of the Competence Center Quantum Computing Baden-W\"urttemberg (project 'QORA').

\appendix

\section{Classical Chernoff bound}
\label{classchernoff}
The classical Chernoff bound gives the rate $\xi$ at which two probability distributions can be distinguished. It is the classical analog of Eq.~\eqref{chernoff}. For two distributions  $g(x)$ and $h(x)$ of a real variable $x\in\mathbb{R}$ it reads
\begin{equation}
\label{eqclasschernoff}
\xi=-\log\inf_{0\leq s\leq 1}\int_{-\infty}^\infty g(x)^{1-s}h(x)^s\,dx.
\end{equation}
The distinguishibility rate in units of the number of channel iterations is $\xi/\Nit$. Applying $\Nit$ times the gate $R_{{\bm e}_z}(\theta)$ to an initial state $\ket{+}$ and then measuring in the $\ket{\pm}$ basis yields outcome $+$ with probability $p_+(\theta)=\cos(\Nit \theta/2)^2$.
If in reality the gate is rather the (fixed) faulty unitary $R_{{\bm e}_z}(\theta')$ with $\theta'=\theta+\varepsilon$, where $\varepsilon$ is a small angle, outcome $+$ is obtained with probability $p_+(\theta')$. The classical Chernoff bound that determines the distinguishability between these two probability distributions of measurement results is obtained from Eq.~\eqref{eqclasschernoff} by minimizing $\sum_{i=\pm}p_i(\theta)^{1-s}p_i(\theta')^{s}$. This sum is equal to
\begin{equation}
 f(s)\equiv x^{1-s}y^s+(1-x)^{1-s}(1-y)^s\,,
 \label{deff}
\end{equation}
where $x=p_+(\theta)$ and $y=p_+(\theta')$.

The function $f$ is convex, and the minimum is reached when $f'(s)=0$. For $x\geq y$ the minimum is at
\begin{equation}
\label{defsmin}
s_{\min}=-\frac{\log \left(\frac{(1-y) \left(\log(1-y)- \log(1-x)\right)}{y (\log x-\log y)}\right)}{\log \left(\frac{(1-x) y}{x (1-y)}\right)},
\end{equation}
and the roles of $x$ and $y$ are exchanged when $y\geq x$. Then Eq.~\eqref{eqclasschernoff} gives $\xi=-\log f(s_{\min})$.

\begin{figure}[!t]
\includegraphics[scale=0.9]{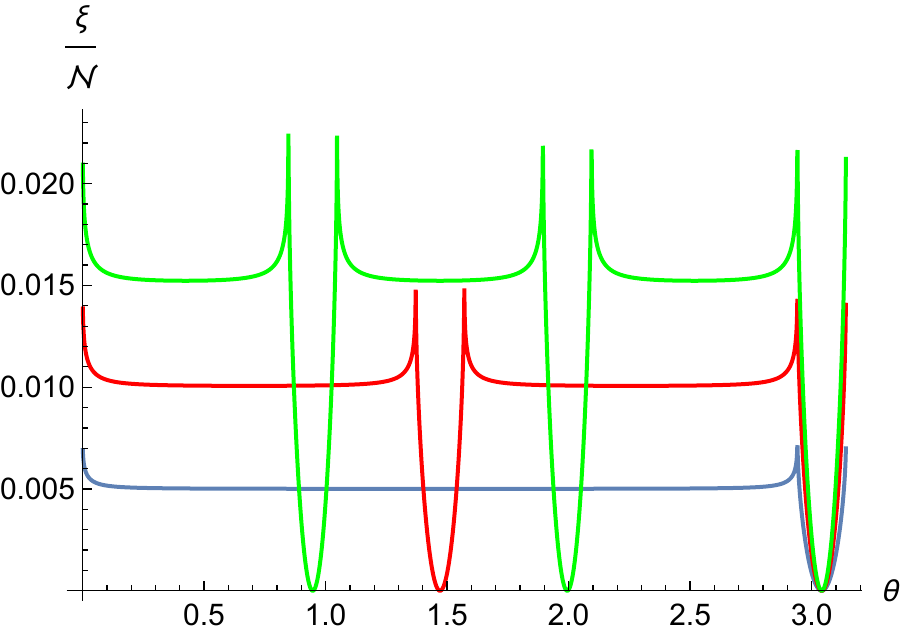}
\caption{Classical Chernoff bound  $\xi=-\log f(s_{\min})$ rescaled by $\Nit$ for the phase gate $R_{{\bm e}_z}(\theta)$, as a
  function of the angle $\theta$ for $\varepsilon=0.2$ and (from bottom
  to top on the left) $\Nit=1$ (blue), 2 (red), 3 (green) iterations of the channel before measurement.
\label{xy}}
\end{figure}

In the present case we have explicitly $x=\cos(\Nit \theta/2)^2$ and $y=\cos(\Nit( \theta+\varepsilon)/2)^2$.  The resulting Chernoff bound is illustrated in Fig.~\ref{xy}. We see that generically iterating the channel is beneficial, e.g.~at small $\theta$, the largest value of $\xi/\Nit$ is associated with the largest $\Nit$.  However, for some values of $\theta$ stopping iterations at $\Nit=2$ or even $\Nit=1$ proves more efficient.
More precisely, the two probability distributions are indistinguishable iff $p_+(\theta)=p_+(\theta +\varepsilon)$,  which implies $\xi=0$. Since $p_+(\theta)=\cos(\mathcal{N}\theta/2)^2$, this occurs at fixed $\varepsilon$ for
\begin{equation}
	\theta_{\Nit,\ell}^*=-\frac{\varepsilon}{2}+\frac{\pi}{\mathcal{N}}\ell,
\end{equation}
with $\ell$ an integer number. At these points $\theta_{\Nit,\ell}^*$, other values of $\Nit,\ell$ fare typically better, unless they also  happen to lead to vanishing $\xi$.
With increasing $\Nit$ the number of zeroes of $\xi$ in the interval $[0,\pi]$ increases, implying more and more such exceptions.

An expansion of $f(s_{\min})$ at small $\varepsilon$, using Eqs.~\eqref{deff}--\eqref{defsmin}, shows that at lowest order $f(s_{\min})\simeq 1-\frac18 \Nit^2 \varepsilon^2$. Then
\begin{equation}
  \label{xiclasscher}
  \xi\simeq \frac18 \Nit^2 \varepsilon^2.
  \end{equation}

\section{Technical details}
\label{technical}
\begin{figure}[!b]
\includegraphics[scale=0.6]{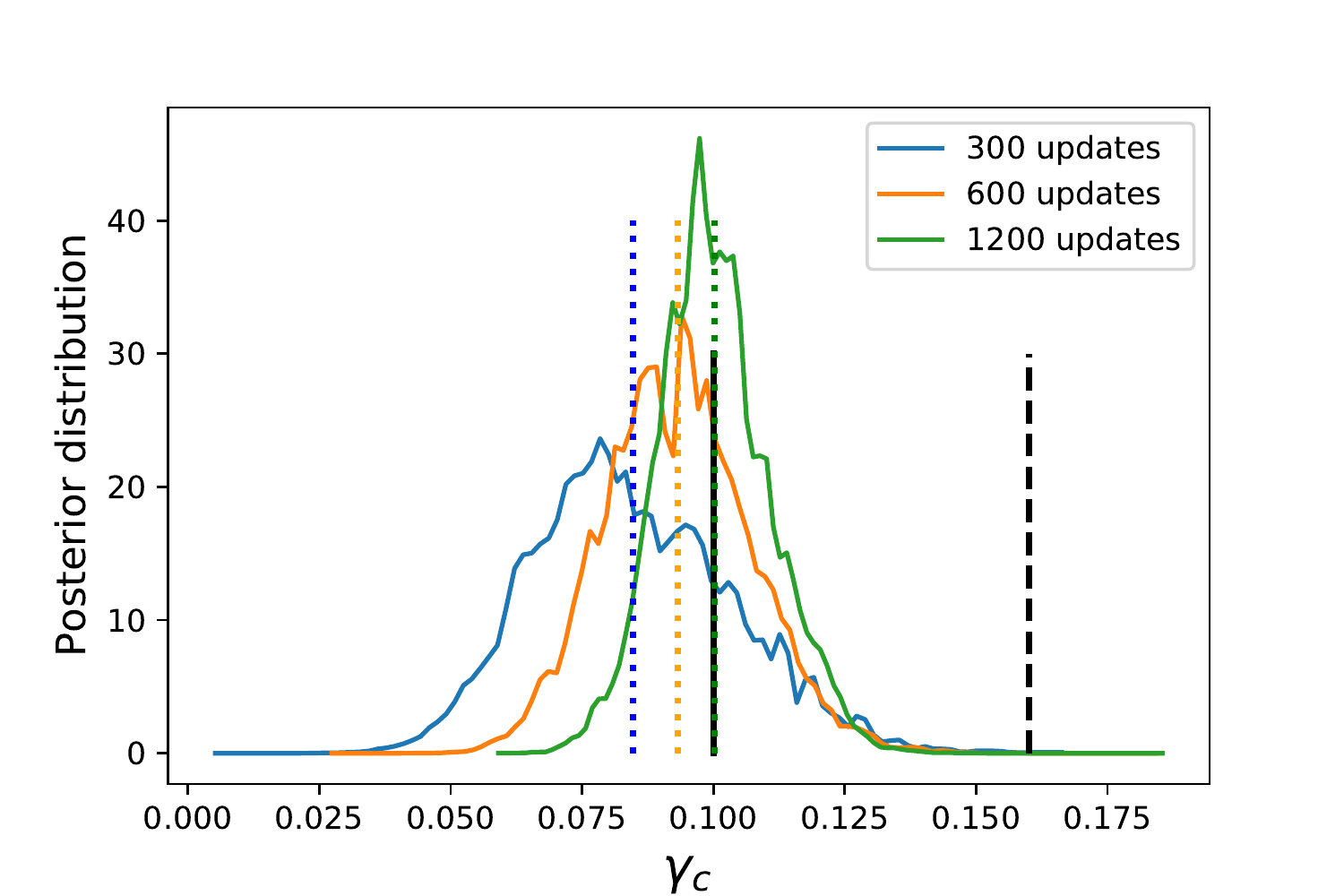}
\caption{Examples of posterior probability distributions for the inference problem of the single parameter $\gamma$ in the case of a phase gate with dephasing for three different number of updates (see legend). Here the solid black line both indicates the true value and the left boundary of the spec, while the dashed black one corresponds to the right boundary of the spec interval. The dotted coloured lines corresponds to the estimated mean of the different posteriors distributions.}
\label{test1}
\end{figure}

In Fig.~\ref{test1} we illustrate the Bayesian approach in the case of a phase gate with dephasing. Starting from a uniform prior distribution $\gamma\in [0,1]$, after a certain number of updates of the probability we end up with posterior distributions which are Gaussian-like. The width of the distributions varies as the square root of the number of updates (see Fig.~\ref{test1}).

\begin{figure}
	\includegraphics[width=.4\textwidth]{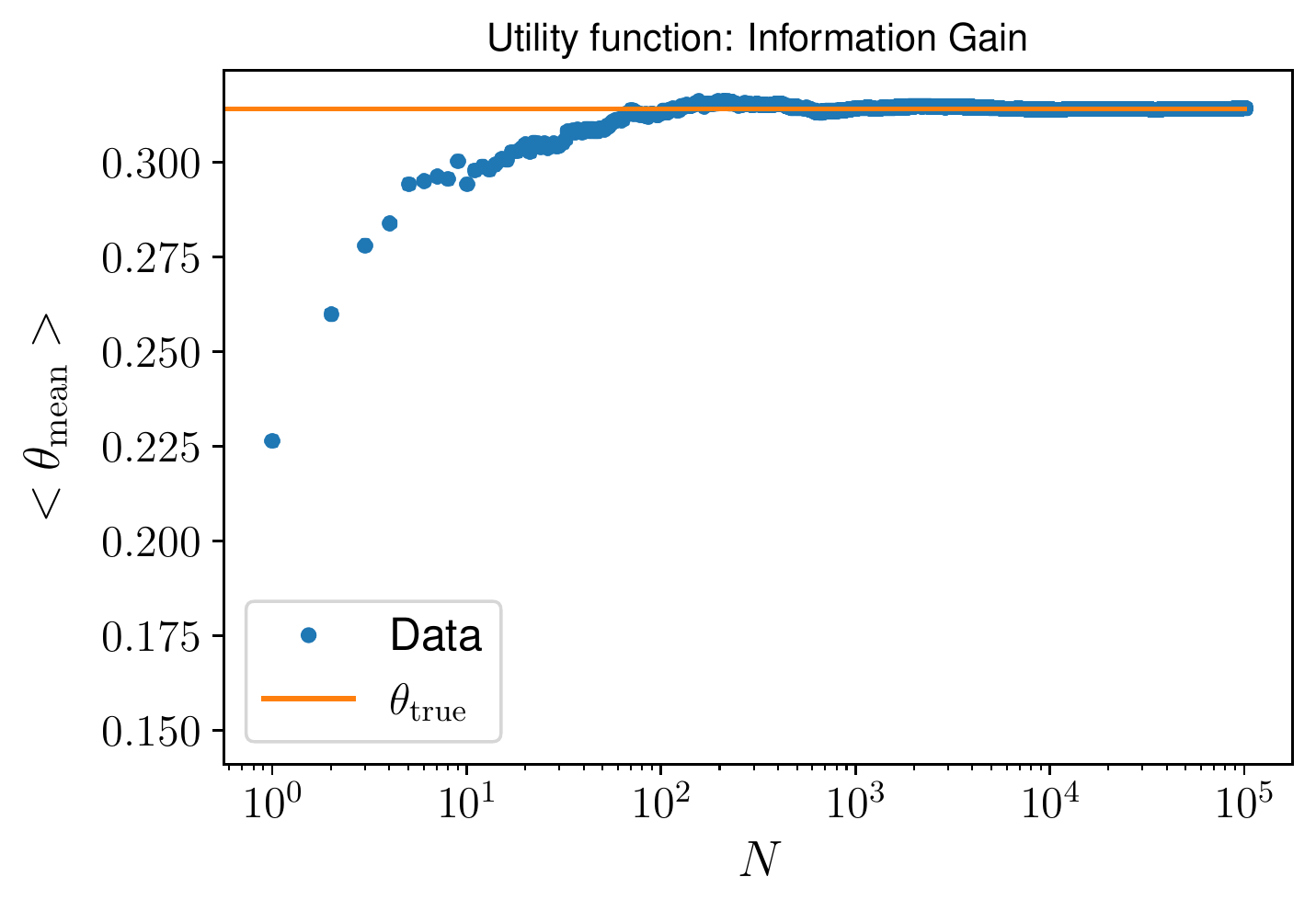}
	\includegraphics[width=.4\textwidth]{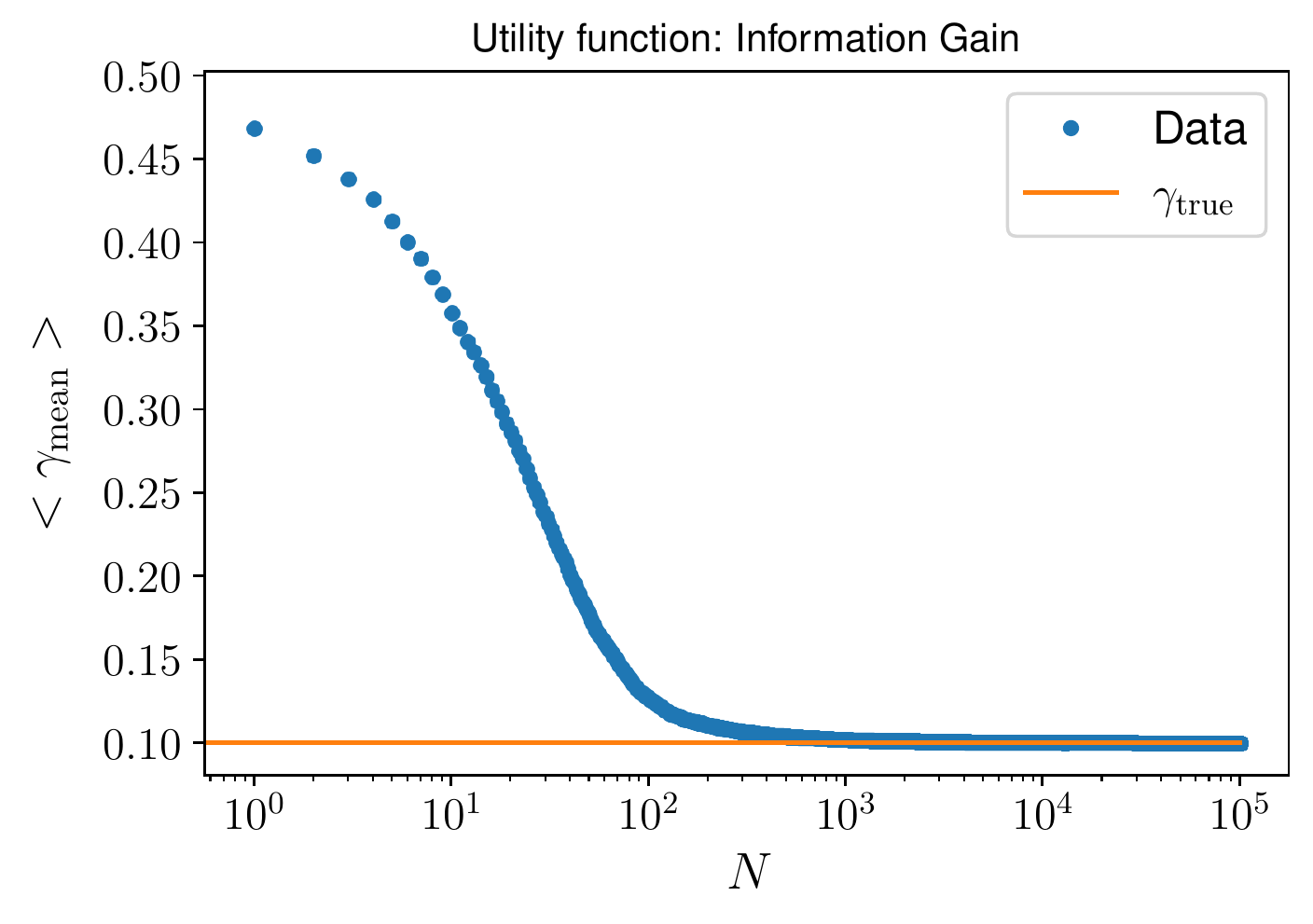}
	\caption{Average value of the mean of the posterior distribution of the parameter (the phase $\theta$ in the upper panel and the decoherence rate $\gamma$ in the lower panel) plotted against the number of updates $N=N_0$ (here $m=1$). The average $\langle\cdot\rangle$ is taken over $2000$ iterations of the estimation procedure.}
	\label{fig:aveMeanEvol}
\end{figure}

\begin{figure}
  \includegraphics[width=.4\textwidth]{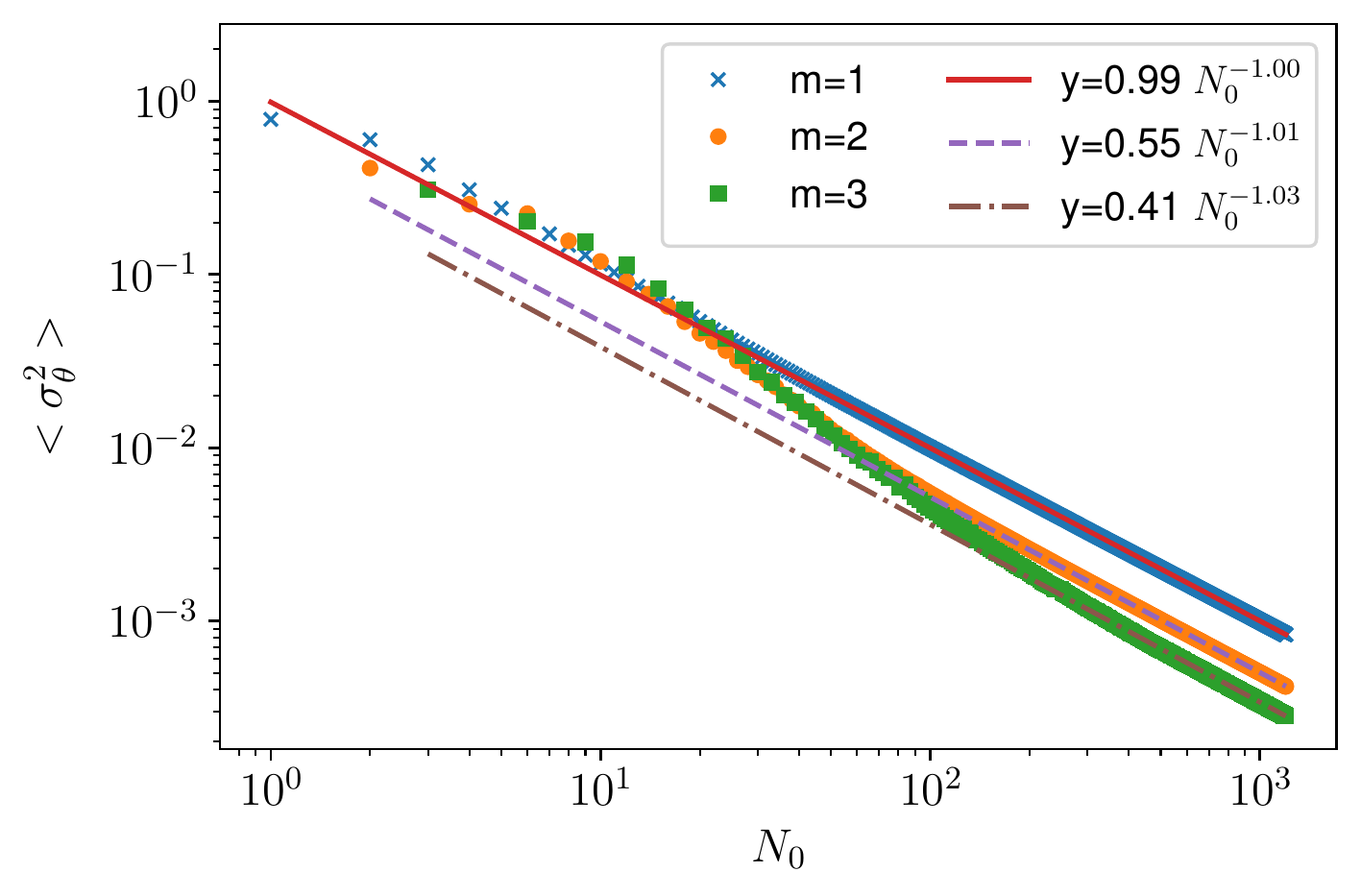}
	\includegraphics[width=.4\textwidth]{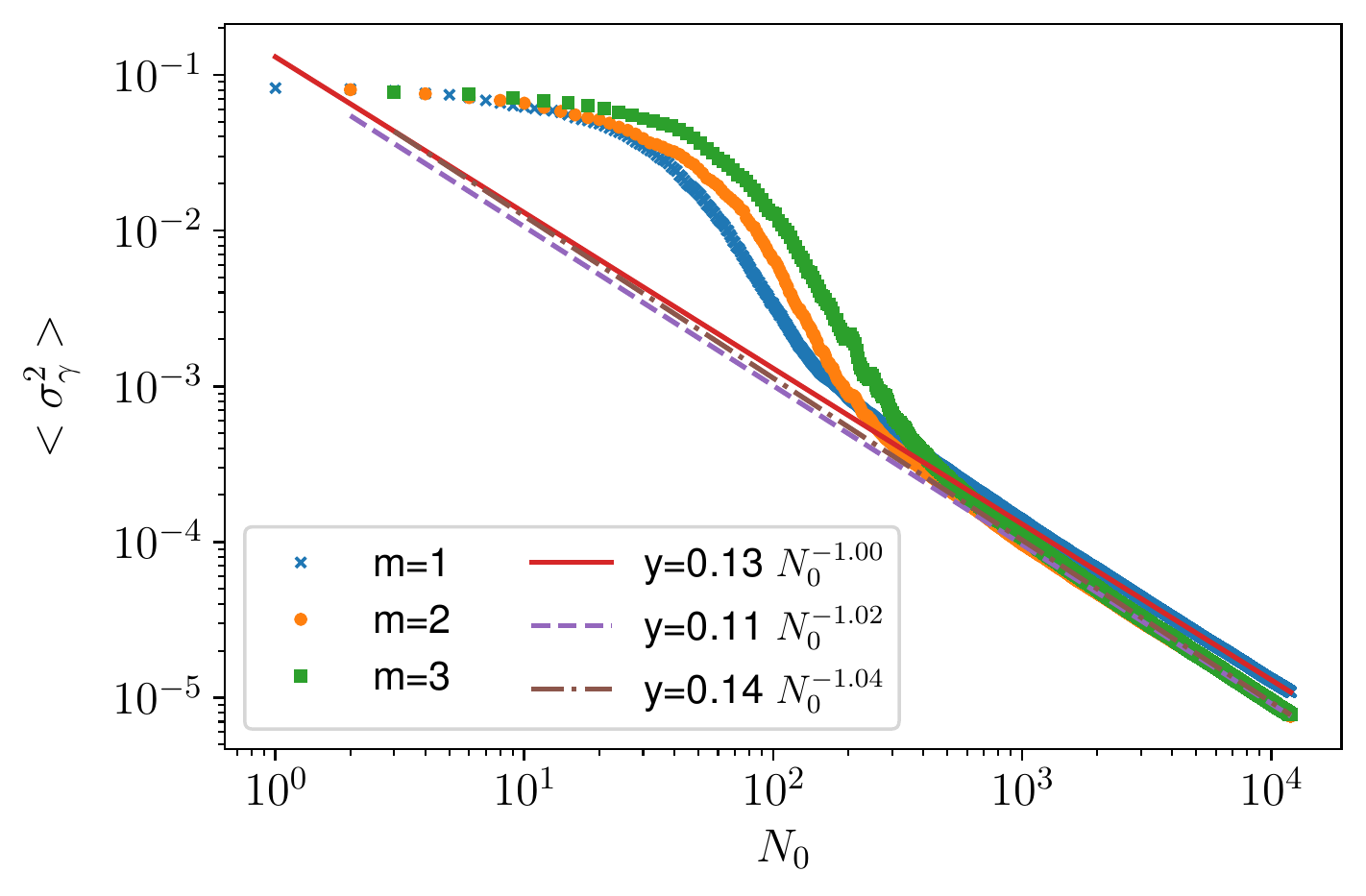}
	\caption{(Upper panel) Average value of the variance of the posterior distribution of the phase $\theta$ with $m=1,2,3$ plotted against the number of total actions $N_0$. The average $\langle\cdot\rangle$ is taken over $140$ iterations of the estimation procedure. The straight lines correspond to  power-law fits taken over the range $[600,1200]$. The power law is always $\simeq 1/N_0$, but with an improvement in the prefactor with larger values of $m$. (Lower panel) As above, but with the distribution of the dephasing rate $\gamma$. In this case the fit is taken over the range $[6000.12000]$. The power law is always $\simeq 1/N_0$, with prefactor quite similar in the three cases $m=1,2,3$, the only difference between the curves being the different time it takes for the onset of the asymptotic regime.}
	\label{fig:aveVarEvol}
\end{figure}

\begin{figure}
  \includegraphics[width=.4\textwidth]{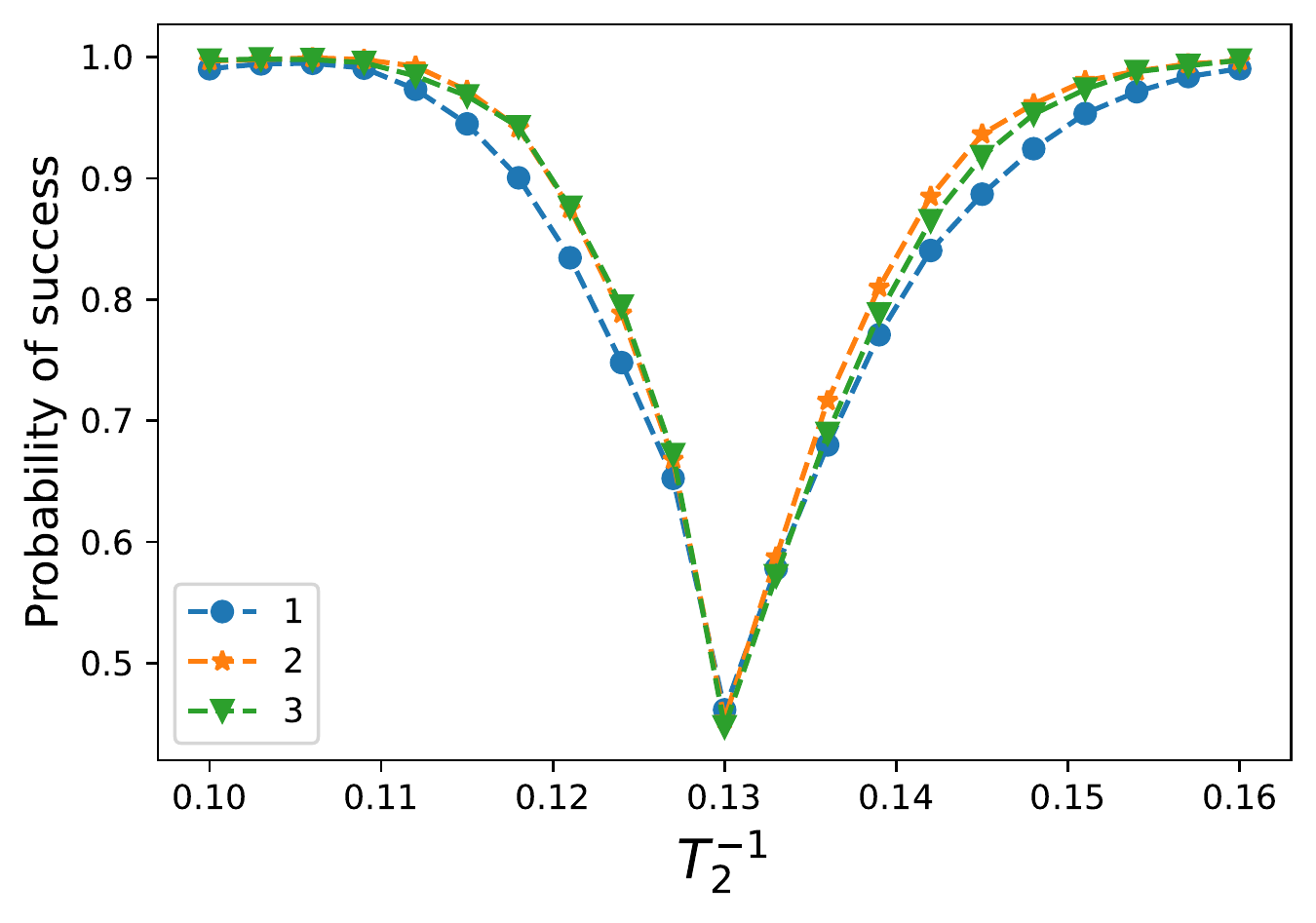}
	\caption{Same as Fig.~\ref{panel2}, but with $N_0=1200$.}
	\label{panel2n01200}
\end{figure}

\figurename{ \ref{fig:aveMeanEvol}} shows the position of the mean of the posterior distribution of the parameters, averaged over $n_{samples}=2000$ iterations of the bayesian estimation and plotted against the number of updates. One can see that after $N\simeq 300$ we are already reasonably close to the true value of the parameter in both cases ($\theta$ for the phase gate and $\gamma$ for the phase gate with dephasing). In particular, in \figurename{ \ref{fig:aveVarEvol}} one can see the behaviour of the variance of the posterior distribution (again averaged over $n_{samples}=2000$ iterations); a fit of the last two decades of points shows that the asymptotic behaviour of the variance is $\sigma_\theta^2\simeq 1/N$ for the phase gate 
with prefactors $0.99$, $0.55$, and $0.41$ for $m=1,2,3$, demonstrating once more the advantage of iterating the channel.  For the phase gate with dephasing, $\sigma_\gamma^2\simeq 0.14/N$. 
Compared to the variance of $\theta$ there is a decrease of the variance of $\gamma$ by a factor of almost $10$ for sufficiently large $\gamma$, which is essentially due to a quite fast decrease in the first $\simeq 100$ steps of the update procedure.
In Fig.~\ref{fig:aveVarEvol} the width for $m=3$ is still larger than for $m=1,2$. This explains why in Fig.~\ref{panel2} the success probability associated with $m=3$ is not always greater than for $m=1,2$.  However, for larger $N_0$ the width for $m=3$ becomes smaller than for $m=1$. In Fig.~\ref{panel2n01200} we show that this translates to success probabilities larger for $m=3$ than for $m=1$ ($m=3$ and $m=2$ are comparable).  

\bibliography{ThesisBib2,mybibs_bt}

\end{document}